\documentclass[journal]{IEEEtran}
\usepackage{amsmath,amssymb,amsthm,lipsum} 
\usepackage{cite}
\usepackage{algorithmic}
\usepackage{algorithm}
\usepackage{graphicx}
\usepackage{textcomp}
\usepackage{stfloats}
\usepackage{booktabs}
\usepackage{bm}
\usepackage{subfigure}
\usepackage{caption}
\usepackage{color}
\ifCLASSINFOpdf
\else
\fi
\theoremstyle{Definition}
\newtheorem{def1}{Definition}
\newtheorem{def2}[def1]{Definition}
\newtheorem{def3}[def1]{Definition}
\newtheorem{def4}[def1]{Definition}

\theoremstyle{Theorem}
\newtheorem{theo}{Theorem}
\newtheorem{theorem1}[theo]{Theorem}
\newtheorem{theorem2}[theo]{Theorem}

\newtheorem{theorem5}[theo]{Theorem}
\newtheorem{theorem6}[theo]{Theorem}
\newtheorem{theorem7}[theo]{Theorem}
\newtheorem{theorem8}[theo]{Theorem}

\theoremstyle{Proposition}
\newtheorem{prop}{Proposition}
\newtheorem{proposition1}[prop]{Proposition}

\theoremstyle{Corollary}
\newtheorem{Corollary1}{Corollary}
\newtheorem{Corollary2}[Corollary1]{Corollary}

\theoremstyle{remark}
\newtheorem{rmk}{Remark}

\newtheorem{rmk3}[rmk]{Remark}
\newtheorem{rmk4}[rmk]{Remark}
\newtheorem{rmk5}[rmk]{Remark}
\newtheorem{rmk6}[rmk]{Remark}
\newtheorem{rmk7}[rmk]{Remark}
\newtheorem{rmk8}[rmk]{Remark}

\theoremstyle{Lemma}
\newtheorem{lemma1}{Lemma}
\newtheorem{lemma2}[lemma1]{Lemma}
\newtheorem{lemma3}[lemma1]{Lemma}
\newtheorem{lemma4}[lemma1]{Lemma}
\newtheorem{lemma5}[lemma1]{Lemma}
\newtheorem{lemma6}[lemma1]{Lemma}
\newtheorem{lemma7}[lemma1]{Lemma}
\newtheorem{lemma8}[lemma1]{Lemma}

\begin{document}

%
\title{Recovery Conditions of Sparse Signals \\
Using Orthogonal Least Squares-Type Algorithms}
%
%
%

\author{Liyang~Lu,~\IEEEmembership{Student Member,~IEEE,}
        Wenbo~Xu,~\IEEEmembership{Member,~IEEE,}
        Yue~Wang,~\IEEEmembership{Member,~IEEE,}
        Zhi~Tian,~\IEEEmembership{Fellow,~IEEE,}
        Yupeng~Cui
        and~Siye~Wang
\thanks{L. Lu, W. Xu, Y. Cui and S. Wang are with the Key Lab of Universal Wireless Communications, Ministry of Education, Beijing University of Posts and Telecommunications.
W. Xu is the corresponding author (e-mail: xuwb@bupt.edu.cn).

Y. Wang and Z. Tian are with the Department of Electrical and Computer Engineering, George Mason University, Fairfax, VA.
}}

%
%

\markboth{}%
{Shell \MakeLowercase{\textit{et al.}}: Bare Demo of IEEEtran.cls for IEEE Journals}
%



\maketitle

\begin{abstract}
Orthogonal least squares (OLS)-type algorithms are efficient in reconstructing sparse signals, which include the well-known OLS, multiple OLS (MOLS) and block OLS (BOLS). In this paper, we first investigate the noiseless exact recovery conditions of these algorithms. Specifically, based on mutual incoherence property (MIP), we provide theoretical analysis of OLS and MOLS to ensure that the correct nonzero support can be selected during the iterative procedure. Nevertheless, theoretical analysis for BOLS utilizes the block-MIP to deal with the block sparsity. Furthermore, the noiseless MIP-based analyses are extended to the noisy scenario. Our results indicate that for $K$-sparse signals, when MIP or SNR satisfies certain conditions, OLS and MOLS obtain reliable reconstruction in at most $K$ iterations, while BOLS succeeds in at most $(K/d)$ iterations where $d$ is the block length. It is shown that our derived theoretical results improve the existing ones, which are verified by
simulation tests.
\end{abstract}


%
\IEEEpeerreviewmaketitle

\section{Introduction}

\IEEEPARstart{I}{n} compressed sensing (CS) \cite{42,70,6}, the main task is to recover a high-dimensional sparse signal from low-dimensional compressed measurements. Specifically, it seeks to accurately recover a $K$-sparse signal $\mathbf{x}\in \mathcal{R}^N$ with $K$ nonzero elements ($K\ll N$) from its linear measurement vector $\mathbf{y}=\mathbf{D}\mathbf{x}$ \cite{51,52,53}, where the measurement matrix $\mathbf{D}$ satisfies restricted isometry property (RIP) \cite{28,43,67}. Exploiting the signal sparsity, a large number of reconstruction algorithms have been developed to exactly recover the original sparse signals under different conditions on $\mathbf{D}$ \cite{2,1,4,9}.

At present, there are two categories of sparse signal recovery approaches: one exploits convex optimization techniques \cite{34,81} and the other is based on greedy matching pursuit \cite{76,77,11,78}. Optimization-based algorithms, such as the well-known basis pursuit (BP) \cite{34}, enjoy satisfactory performance, at the cost of high computational complexity that may not be amenable to practical implementation. On the other hand, greedy algorithms, such as matching pursuit (MP) and orthogonal MP (OMP) \cite{90,91}, admit simple and fast implementations, and hence have been widely used in practice. Another popular family of greedy algorithms is the orthogonal least squares (OLS) type, which is shown to have preferable convergence property \cite{14}. This paper focuses on the family of OLS-type algorithms for sparse signal recovery, because of their computational simplicity and competitive performance.

OLS is an iterative CS algorithm that picks out the support of the sparse signal by updating only one index to the list per iteration. Then, several variants are developed to enhance the sparse recovery performance. In \cite{9}, multiple OLS (MOLS) is proposed which selects multiple indices per iteration. In \cite{80}, a block version of the OLS algorithm, named BOLS, is proposed to utilize the block structure of the underlying signals. It works well for practical scenarios where the nonzero support of the sparse signal appears in clusters rather than spreading out randomly \cite{64,65}.

A fundamental problem in CS reconstruction is to characterize the recovery capabilities of these algorithms, where RIP is usually used as a representative metric. The authors in \cite{4,9,88} exploit RIP to analyze the isometry constant of OLS for exactly recovering a $K$-sparse signal in $K$ iterations. In \cite{9}, the authors utilize RIP to obtain the recovery conditions for MOLS and show its performance in noisy case. Furthermore, \cite{10} uses similar methodology to analyze MOLS and proposes a less restrictive performance guarantee condition.

Unfortunately, to obtain the RIP constant of a given measurement matrix is NP-hard. By contrary, a computationally friendly approach is the mutual incoherence property (MIP) \cite{2,58,60}. This technique has been utilized to analyze the performance guarantee of OMP-type algorithms in the last decade. Compared with the OMP-type, the OLS-type algorithms are more reliable and the performance is less dependent on the amplitude distribution of nonzero entries \cite{89}. However, the analysis of OLS-type algorithms is more involved, because their atom selection rules are more complicated than those of the OMP-type algorithms. As a result, the study on OLS-type algorithms with MIP tool is seldom in current literature. Though the authors in \cite{89} present the MIP-based conditions of OLS, the results only apply for decaying signals. To better understand their recovery capabilities for general sparse signals and show comparison with OMP-type algorithms \cite{2,1}, it is necessary to investigate the MIP-based performance for the OLS-type algorithms.

Most work on MIP-based analysis is devoted to the recovery of noiseless signals \cite{2,1}. Nevertheless, noise is always present in practical applications \cite{89,85}. A fundamental question is thus at what signal-to-noise ratio (SNR) levels the OLS-type algorithms realize reliable recovery. Since the MIP metric is powerful in assessing  recovery performance, MIP analysis can shed light on the target SNR levels for ensuring signal reconstruction at high accuracy.

Based on the above observations, this paper focuses on the recovery conditions for OLS-type algorithms, both in the noiseless and the noisy scenarios. Note that the mapping normalization factor in OLS-type algorithms makes the analysis greatly different from that of the OMP-type in \cite{2,1}, which poses major challenges \cite{4,14}. We shall show that the MIP technique is effective to bound this mapping normalization factor.
Another major difference from the study in \cite{2,1} is that we consider the noise effect. We develop the MIP and SNR-based recovery conditions for correct atom selection when additive noise is present. It is shown that the OLS-type algorithms can choose all the correct atoms from the measurement matrix under these conditions and thus guarantee their performance. The contributions of this paper are summarized as below.

\begin{enumerate}[]

    \item For the noiseless case, we develop the MIP-based asymptotic exact recovery conditions (ERCs) of the OLS-type algorithms. The derived conditions imply higher reconstructible sparsity levels than those in \cite{2,1}. Our analytical results reveal that the OLS-type algorithms can recover any $K$-sparse or block $k$-sparse signal within $K$ or $k$ iterations under the derived conditions, respectively.

    \item In the noisy case, the recovery conditions are developed for OLS-type algorithms by extending the noiseless results. The proposed noisy recovery conditions imply lower signal power required for reliable recovery compared to the known bound in \cite{85}. These results unveil that if the nonzero coefficients of a sparse signal are larger than the derived bounds specified by the MIP and noise variance, the OLS-type algorithms can achieve reliable recovery.

    \item Validating simulations are performed to confirm the derived theoretical results. It is shown that the theoretical guarantees in this paper lead to better bounds compared with the existing ones. Empirical simulations indicate that under the derived conditions, the OLS-type algorithms can reliably reconstruct the original signals.

\end{enumerate}

The rest of this paper is organized as follows. In Section~${\rm\uppercase\expandafter{\romannumeral2}}$, we introduce notations, CS background and OLS-type algorithms. In Section ${\rm \uppercase\expandafter{\romannumeral3}}$, we study the noiseless recovery conditions for OLS-type algorithms. In Section ${\rm \uppercase\expandafter{\romannumeral4}}$, the noisy recovery conditions for OLS-type algorithms  are derived. In Section ${\rm \uppercase\expandafter{\romannumeral5}}$, simulation tests are presented, followed by conclusions in Section ${\rm \uppercase\expandafter{\romannumeral6}}$.

\section{Preliminaries}
\subsection{Notations}
In this paper, we denote vectors by boldface lowercase letters, e.g., $\mathbf{r}$, and matrices by boldface uppercase letters, e.g., $\mathbf{D}$. The $i$-th element of $\mathbf{r}$ is denoted as $\mathbf{r}_i$. The element of matrix $\mathbf{D}$ is denoted as $\mathbf{D}_{ij}$ and $\mathbf{D}_i$ is the $i$-th column of $\mathbf{D}$. $\mathbf{D}^T$ represents the transpose of matrix $\mathbf{D}$. $<\mathbf{D}_i,\mathbf{D}_j>$ is the inner product of $\mathbf{D}_i$ and $\mathbf{D}_j$. Given a constant $c$, $|c|$ means its absolute value.
$\mathbf{D}_{\mathbf{S}}$ is a sub matrix of $\mathbf{D}$ that contains columns indexed by $\mathbf{S}$. $\mathbf{D}\backslash \mathbf{D}_{\mathbf{S}}$ is the set of the entries contained in $\mathbf{D}$ but not in $\mathbf{D}_{\mathbf{S}}$.
Suppose $\mathbf{D}_{\mathbf{S}}$ has full column rank, $\mathbf{P}_{\mathbf{S}}=\mathbf{D}_{\mathbf{S}}\mathbf{D}_{\mathbf{S}}^\dag$ stands for the projection onto ${\rm span}(\mathbf{D}_{\mathbf{S}})$, where $\mathbf{D}_{\mathbf{S}}^\dag=(\mathbf{D}_{\mathbf{S}}^T\mathbf{D}_{\mathbf{S}})^{-1}\mathbf{D}^T_{\mathbf{S}}$ is the pseudoinverse of $\mathbf{D}_{\mathbf{S}}$. $\mathbf{P}_{\mathbf{S}}^\bot=\mathbf{I}-\mathbf{P}_{\mathbf{S}}$ is the projection onto the orthogonal complement of ${\rm span}(\mathbf{D}_{\mathbf{S}})$. The $\ell_0$-norm, $\ell_1$-norm, $\ell_2$-norm, $\ell_\infty$-norm and $\ell_{2,\infty}$-norm of $\mathbf{r}$ are represented by $||\mathbf{r}||_0$, $||\mathbf{r}||_1$, $||\mathbf{r}||_2$, $||\mathbf{r}||_\infty$ and $||\mathbf{r}||_{2,\infty}$ respectively. The spectral norm of $\mathbf{D}$ is denoted by $\rho(\mathbf{D})=\sqrt{\lambda_{\max}(\mathbf{D}^T\mathbf{D})}$, where $\lambda_{\max}(\cdot)$ is the largest eigenvalue of its objective. The set consists of the indices of the nonzero elements in $\mathbf{r}$ is represented by ${\rm supp}(\mathbf{r})$. The diagonalization of $\mathbf{r}$ is denoted by ${\rm diag}(\mathbf{r})$.

\subsection{Compressed sensing and sparse signal models}
The basic model of CS is given as follows:
\begin{equation}\label{CSmodel}
\mathbf{y}=\mathbf{D}\mathbf{x},
\end{equation}
where $\mathbf{y}\in \mathcal{R}^{M}$ is the low-dimensional measurement vector, $\mathbf{D}\in \mathcal{R}^{M\times N}$ is the measurement matrix and $\mathbf{x}\in \mathcal{R}^{N}$ is the signal vector with $N>M$. Generally, infinitely many choices of $\mathbf{x}$ satisfy (\ref{CSmodel}) for a given $\mathbf{y}$ because (\ref{CSmodel}) is underdetermined. Therefore, in order to guarantee uniqueness of $\mathbf{x}$, CS considers the case of sparse signals, i.e., there are only a few nonzero elements in $\mathbf{x}$ relative to its dimension. The problem of obtaining the sparse solution to (\ref{CSmodel}) can be expressed as
\begin{equation}\label{l0}
\min_{\mathbf{x}}||\mathbf{x}||_0\quad {\rm s.t.} \quad\mathbf{y}=\mathbf{D}\mathbf{x}.
\end{equation}
This $\ell_0$-norm minimization problem is non-convex and NP-hard. Fortunately, the problem can be relaxed by its convex surrogate $\ell_1$ optimization \cite{70},
\begin{equation}\label{l1}
\min_{\mathbf{x}}||\mathbf{x}||_1\quad {\rm s.t.} \quad\mathbf{y}=\mathbf{D}\mathbf{x}.
\end{equation}
Many methods mentioned in the first section can be applied to solve (\ref{l1}).

When $\mathbf{x}$ is block sparse, block CS is proposed in \cite{1,73,74,75} to recover the signal. Letting $d$ denote the block length, the block sparse signal $\mathbf{x}$ is defined as
\begin{equation}\label{sparsex}
\mathbf{x}=[\underbrace{\mathbf{x}_1\cdots  \mathbf{x}_d }_{\mathbf{x}^T[1]} \underbrace{\mathbf{x}_{d+1}\cdots \mathbf{x}_{2d}}_{\mathbf{x}^T[2]}\cdots \underbrace{\mathbf{x}_{N-d+1}\cdots \mathbf{x}_{N}}_{\mathbf{x}^T[N_B]}]^T,
\end{equation}
where $N=N_Bd$ and $\mathbf{x}[i]\in \mathcal{R}^{d}$ is the $i$-th block of $\mathbf{x}$ $(i\in\{1,2,\cdots,N_B\})$. A signal $\mathbf{x}$ is called block $k$-sparse if $\mathbf{x}$ has $k$ nonzero $\ell_2$-norm blocks. The measurement matrix can be rewritten as a concatenation of $N_B$ column blocks, i.e.,
\begin{equation}\label{matrixblock}
\mathbf{D}=[\underbrace{\mathbf{D}_1\cdots  \mathbf{D}_d }_{\mathbf{D}[1]} \underbrace{\mathbf{D}_{d+1}\cdots \mathbf{D}_{2d}}_{\mathbf{D}[2]}\cdots \underbrace{\mathbf{D}_{N-d+1}\cdots \mathbf{D}_{N}}_{\mathbf{D}[N_B]}],
\end{equation}
where $\mathbf{D}[i]\in \mathcal{R}^{M\times d}$ is the $i$-th block of $\mathbf{D}$. The next subsection describes the OLS-type algorithms for reconstructing the sparse $\mathbf{x}$ from $\mathbf{y}$ in (\ref{CSmodel}).

\subsection{OLS-type algorithms}

\subsubsection{OLS Algorithm}
\begin{algorithm}
	\renewcommand{\algorithmicrequire}{\textbf{Input:}}
	\renewcommand{\algorithmicensure}{\textbf{Output:}}
	\caption{Orthogonal least squares}
	\label{alg:11}
	\begin{algorithmic}[1]
		\REQUIRE $\mathbf{D}, \mathbf{y}$, total sparsity level $K$ and residual tolerant $\zeta$
		\ENSURE $\mathbf{x}\in\mathcal{R}^{N}$, $\mathbf{S}\subseteq \{1,2,\cdots,N\}$
        \STATE $\mathbf{Initialization:}$ $l=0$, $\mathbf{r}^0=\mathbf{y}$, $\mathbf{S}^0=\emptyset$, $\mathbf{x}^0=\mathbf{0}$
        \WHILE {$l< K$ and $||\mathbf{r}^l||_2>\zeta$}
		\STATE Set $i^{l+1}=\mathop{\arg\min}\limits_{j\in\{1,\cdots,N\}\backslash\mathbf{S}^{l}}||\mathbf{P}^\bot_{\mathbf{S}^{l}\cup \{j\}}\mathbf{y}||_2^2$
		\STATE Augment $\mathbf{S}^{l+1}=\mathbf{S}^{l}\cup{\{i^{l+1}\}}$
		\STATE Estimate $\mathbf{x}^{l+1}=\mathop{\arg\min}\limits_{\mathbf{x}: \;{\rm supp}(\mathbf{x})=\mathbf{S}^{l+1}}\|\mathbf{y}-\mathbf{D}\mathbf{x}\|_2^2$
        \STATE Update $\mathbf{r}^{l+1}=\mathbf{y}-\mathbf{D}\mathbf{x}^{l+1}$
		\STATE $l=l+1$
        \ENDWHILE
		\STATE \textbf{return} $\mathbf{S}=\mathbf{S}^{l}$ and $\mathbf{x}=\mathbf{x}^{l}$
	\end{algorithmic}
\end{algorithm}
The standard OLS algorithm is given in Algorithm \ref{alg:11}.
To facilitate the theoretical derivations in ensuing sections, we rewrite Step~3 in Algorithm 1 in an alternative form:
\begin{equation}\label{alterformula}
\begin{aligned}
i^{l+1}&=\mathop{\arg\max}\limits_{j\in\{1,..,N\}\backslash\mathbf{S}^{l}}\Bigg|\Bigg<\frac{\mathbf{P}^\bot_{\mathbf{S}^{l}}\mathbf{D}_j}{||\mathbf{P}^\bot_{\mathbf{S}^{l}}\mathbf{D}_j||_2},\mathbf{r}^l\Bigg>\Bigg|\\
&=\mathop{\arg\max}\limits_{j\in\{1,..,N\}\backslash\mathbf{S}^{l}}\Bigg|\Bigg<\frac{\mathbf{D}_j}{||\mathbf{P}^\bot_{\mathbf{S}^{l}}\mathbf{D}_j||_2},\mathbf{r}^l\Bigg>\Bigg|.
\end{aligned}
\end{equation}
The derivation details can be found in \cite{9}\cite{20}. This formula also offers a more intuitive explanation of OLS. Specifically, it selects a candidate column that is most closely related with the current residual after the column is orthogonally projected onto the subspace ${\rm span}(\mathbf{D}_{\mathbf{S}^l})$.

\subsubsection{MOLS Algorithm}
In the $l$-th iteration, the support selection criteria of MOLS algorithm \cite{9} is given as follows:
\begin{equation}\label{mols select}
\mathbf{Q}^{l+1}_M=\mathop{\arg\min}\limits_{\mathbf{Q}_M:\;|\mathbf{Q}_M|=L}\sum_{j_M\in\mathbf{Q}_M}||\mathbf{P}^\bot_{\mathbf{S}^{l}\cup \{j_M\}}\mathbf{y}||_2^2.
\end{equation}
Owning to the selection of multiple candidates per iteration, MOLS converges faster than OLS.

\subsubsection{BOLS Algorithm}
When $\mathbf{x}$ is block sparse, \cite{80} proposes a tailored BOLS algorithm, which seeks a block candidate that makes the most significant reduction in the residual power. The support selection criteria of BOLS is given by
\begin{equation}\label{bolsselect}
i^{l+1}_B=\mathop{\arg\min}\limits_{j_B\in\{1,\cdots,N_B\}\backslash \mathbf{S}_B^{l}}||\mathbf{P}^\bot_{\mathbf{S}^{l}\cup \{(j_B-1)d+1,\cdots,j_Bd\}}\mathbf{y}||_2^2,
\end{equation}
where the entries in $\mathbf{S}_B^{l}$ correspond to the block indices of $\mathbf{S}^{l}$.
All these algorithms follow the same procedure, except that the support selection criteria are different.

\section{Exact Recovery Conditions}
\label{section3}

\subsection{ERCs for OLS and MOLS algorithms}

In the absence of block sparsity, we provide the ERCs for OLS and MOLS. Relevant concepts are defined below.

\begin{def1}
(Matrix coherence \cite{2}\cite{1}) The coherence of a matrix $\mathbf{D}$, which represents the similarity of its elements, is defined as
\begin{equation}\label{coherence}
\mu=\max_{i,j\neq i}|<\mathbf{D}_i,\mathbf{D}_j>|.
\end{equation}
\label{definitionofcoherence}
\end{def1}

\begin{def2}
(Mixed norm \cite{2,1}) Given a matrix $\mathbf{D}\in \mathcal{R}^{M\times N}$, $||\mathbf{D}||_{1,1}$ equals the maximum absolute column sum of $\mathbf{D}$, i.e.,
\begin{equation}\label{mixednorm}
||\mathbf{D}||_{1,1}=\max_{j}\sum_{i=1}^{M}|\mathbf{D}_{ij}|,
\end{equation}
and $||\mathbf{D}||_{\infty,\infty}$ represents the maximum absolute row sum of $\mathbf{D}$, i.e.,
\begin{equation}\label{mixednorm2}
||\mathbf{D}||_{\infty,\infty}=\max_{i}\sum_{j=1}^{N}|\mathbf{D}_{ij}|.
\end{equation}
Let $M=md$ and $N=N_Bd$,
\begin{equation}\label{rou}
\rho_c(\mathbf{D})=\max_j\sum_i\rho(\mathbf{D}[i,j])
\end{equation}
and
\begin{equation}\label{rou}
\rho_r(\mathbf{D})=\max_i\sum_j\rho(\mathbf{D}[i,j]),
\end{equation}
where $m$ is an integer and $\mathbf{D}[i,j]$ is the $(i,j)$-th block of $\mathbf{D}$.
\label{mixednorm}
\end{def2}
Without loss of generality, suppose that the first $K$ elements of $\mathbf{x}$ are nonzero.
With the set of the selected indices, i.e., $\mathbf{S}^l=\{K-l+1,\cdots,K\}$, in the $(l+1)$-th iteration, we define
\begin{equation}\label{q0explain}
\mathbf{Q}_0=[\mathbf{D}_1,\mathbf{D}_2,\cdots,\mathbf{D}_K]\in \mathcal{R}^{M\times K},
\end{equation}
\begin{equation}\label{q0explain2s}
\mathbf{Q}_{0\backslash\mathbf{S}}=\mathbf{Q}_0\backslash\mathbf{Q}_{\mathbf{S}^l}\in \mathcal{R}^{M\times (K-|\mathbf{S}^l|)}
\end{equation}
and
\begin{equation}\label{qofei}
\overline{\mathbf{Q}}_0=\mathbf{D}\backslash\mathbf{Q}_0\in \mathcal{R}^{M\times (N-K)}.
\end{equation}
$\mathbf{R}_{0\backslash\mathbf{S}}$ and $\overline{\mathbf{R}}_0$ corresponding to $\mathbf{Q}_{0\backslash\mathbf{S}}$ and $\overline{\mathbf{Q}}_0$ are defined by
\begin{equation}\label{Rodefinition}
\begin{aligned}
\mathbf{R}_{0\backslash\mathbf{S}}= f(\mathbf{Q}_{0\backslash\mathbf{S}})
=\left[
\begin{matrix}

   \frac{1}{||\mathbf{P}^\bot_{\mathbf{S}^{l}}\mathbf{D}_1||_2} &  &\mathbf{0}\\

     &       \ddots&\\
     \mathbf{0}&&\frac{1}{||\mathbf{P}^\bot_{\mathbf{S}^{l}}\mathbf{D}_{K-l}||_2}
\end{matrix}
\right]
\end{aligned}
\end{equation}
and $\overline{\mathbf{R}}_0= f(\overline{\mathbf{Q}}_0)$.
Using these arguments, we present the ERCs for OLS and MOLS in the following theorems.
\begin{theorem1}
Assume that OLS or MOLS has chosen $l$ correct atoms after $l$ iterations. A sufficient condition for OLS to select one correct atom or MOLS to select at least one correct atom in the $(l+1)$-th iteration is
\begin{equation}\label{olsksteps}
||(\mathbf{Q}_{0\backslash\mathbf{S}}\mathbf{R}_{0\backslash\mathbf{S}})^\dag(\overline{\mathbf{Q}}_{0}\overline{\mathbf{R}}_{0})||_{1,1}<1.
\end{equation}
\label{theorem1}
\end{theorem1}
\begin{IEEEproof}
See Appendix \ref{proofoftheorem1}.
\end{IEEEproof}

Then, we present the MIP-based sufficient condition to ensure that Theorem \ref{theorem1} can be established in each iteration.
To obtain this sufficient condition, we give the subsequent lemmas. The first one raises the bounds on the mapping normalization factor, i.e., $||\mathbf{P}^{\bot}_{\mathbf{S}^l}\mathbf{D}_i||_2$, in terms of the MIP.
The second one presents the probability of a random event.

\begin{lemma3}
Suppose $\mu<\frac{1}{K-1}$, then $\frac{1}{\sqrt{\mathcal{T}}}\leq||\mathbf{P}^{\bot}_{\mathbf{S}^l}\mathbf{D}_i||_2\leq1$ for $i\in\{1,2,\cdots,N\}\backslash \mathbf{S}^l$,
where $\mathcal{T}=\big(1-\frac{(1+(K-1)\mu)K\mu^2}{(1-(K-1)\mu)^2}\big)^{-1}$.
\label{lemma3}
\end{lemma3}

\begin{IEEEproof}
See Appendix B.
\end{IEEEproof}

Notably, under the same assumption made as in Lemma \ref{lemma3}, i.e., $\mu<\frac{1}{K-1}$, the following bounds can be obtained by direct calculations using the Lemmas 2 and 5 in \cite{85}:
\begin{equation}\label{lemma2and5}
\sqrt{1-K\mu}\leq||\mathbf{P}^{\bot}_{\mathbf{S}^l}\mathbf{D}_i||_2\leq1.
\end{equation}
It is worthy of noting that $\frac{1}{\sqrt{\mathcal{T}}}\geq\sqrt{1-K\mu}$ since $\mu<\frac{1}{K-1}$, which means Lemma \ref{lemma3} improves the result (\ref{lemma2and5}).
\begin{lemma4}
For a random matrix $\mathbf{B}\in \mathcal{R}^{M\times K}$ $(K\geq2)$ whose entries are independently and identically distributed as $\mathcal{N}(0,\frac{1}{M})$, the following probability holds:
\begin{equation}\label{probability}
{\rm P}\Big\{||\mathbf{B}^T\mathbf{B}-\mathbf{I}||_{1,1}\leq \frac{(K-\mathcal{T})\tau}{2}\Big\}\geq1-\frac{Ke^{-\frac{K-1}{2}(\mathcal{C}-\log(1+\mathcal{C}))}}{\mathcal{C}\sqrt{\pi(K-1)}},
\end{equation}
where $\tau$ is a positive constant and $\mathcal{C}=\frac{M(K-\mathcal{T})^2\tau^2}{4(K-1)^2}-1$.
\label{lemma4}
\end{lemma4}
\begin{IEEEproof}
See Appendix C.
\end{IEEEproof}
\begin{rmk6}
Suppose that the matrix $\mathbf{B}$ in Lemma \ref{lemma4} is a sub matrix of the Gaussian matrix $\mathbf{H}\in \mathcal{R}^{M\times N}$ $(N>M)$. Since $\mu<\frac{1}{K-1}$ and $\mathcal{T}$, which is given in Lemma \ref{lemma3}, is monotonically increasing with respect to $\mu$, then $\mathcal{T}$ is limited by an upper bound. When $K$ is fixed, we obtain
\begin{equation}\label{probis1}
\lim_{M\rightarrow\infty} \frac{M(K-\mathcal{T})^2\tau^2}{4(K-1)^2}-1\rightarrow\infty.
\end{equation}
Therefore,
\begin{equation}\label{probis2}
\lim_{M\rightarrow\infty} 1-\frac{Ke^{-\frac{K-1}{2}(\mathcal{C}-\log(1+\mathcal{C}))}}{\mathcal{C}\sqrt{\pi(K-1)}}=1.
\end{equation}
\label{rmk6}
\end{rmk6}

\begin{theorem2}
Let $\mu$ be the matrix coherence of the measurement matrix $\mathbf{D}$. The ERC (\ref{olsksteps}) is satisfied if
\begin{equation}\label{theorem2mian2}
K<\Big(\sqrt[3]{-\frac{q}{2}+\sqrt{\Delta}}+\sqrt[3]{-\frac{q}{2}-\sqrt{\Delta}}-\frac{\beta}{3\alpha}\Big),
\end{equation}
where $q=\frac{27\alpha^2\delta-9\alpha\beta\gamma+2\beta^3}{27\alpha^3}$, $p=\frac{3\alpha\gamma-\beta^2}{3\alpha^2}$, $\alpha=-\frac{\mu^4}{2}+\frac{3}{2}\mu^3$, $\beta=\frac{\mu^4}{2}-3\mu^3-4\mu^2$, $\gamma=\frac{3\mu^3}{2}+7\mu^2+\frac{7}{2}\mu$, $\delta=-\frac{1}{2}\mu^3-2\mu^2-\frac{5}{2}\mu-1$ and $\Delta=(\frac{q}{2})^2+(\frac{p}{3})^3$.
\label{theorem2}
\end{theorem2}

\begin{IEEEproof}
See Appendix \ref{proofoftheorem2}.
\end{IEEEproof}

\begin{rmk}
Denote the right-side of (\ref{theorem2mian2}) as $C(\mu)$. The upper bound of $C(\mu)$ is $C(\sqrt{\frac{N-M}{M(N-1)}})$, where $M$ and $N$ are the dimensions of the measurement matrix. Moreover, we have the following observations.
(1) When $M$ is fixed, the upper bound of $C(\mu)$ degrades with the increase of $N$.
(2) When $N$ is fixed, the upper bound of $C(\mu)$ is improved with the increase of $M$.
(3) The upper bound is directly related to the number of measurements $M$ and the compression ratio $\omega=\frac{M}{N}$. That is, when $N\gg0$ and $M\gg0$, it holds that
\begin{equation}\label{uppperboundofC1}
C(\mu)\leq\frac{2}{3}\bigg(\sqrt{\frac{M}{1-\omega}}+1\bigg).
\end{equation}
\label{rmk}
\end{rmk}

\begin{IEEEproof}
It is known that $\mu\geq\sqrt{\frac{N-M}{M(N-1)}}$ \cite{82}. Meanwhile, the partial derivative of $C(\mu)$ with respect to $\mu$ satisfies $\nabla C(\mu)<0$. Therefore, $C(\mu)\leq C\big(\sqrt{\frac{N-M}{M(N-1)}}\big)$.

When $M$ is fixed, $\sqrt{\frac{N-M}{M(N-1)}}=\frac{1}{\sqrt{M}}\sqrt{1-\frac{M-1}{N-1}}$ is improved with the increase of $N$, leading to the degradation of $C\big(\sqrt{\frac{N-M}{M(N-1)}}\big)$.
When $N$ is fixed, $\sqrt{\frac{N-M}{M(N-1)}}=\sqrt{\frac{\frac{N}{M}-1}{N-1}}$ decreases with the increase of $M$, which causes the improvement of $C\big(\sqrt{\frac{N-M}{M(N-1)}}\big)$.

Moreover, in asymptotic case, we have
\begin{align}
\lim_{M/N=\omega;\;M,N\rightarrow\infty} C\bigg(\sqrt{\frac{N-M}{M(N-1)}}\bigg)=&C\bigg(\sqrt{\frac{1-\frac{M}{N}}{M(1-\frac{1}{N})}}\bigg)\nonumber\\
=& C\Big(\sqrt{\frac{1-\omega}{M}}\Big).\label{nmrelation}
\end{align}

Rewrite $C\big(\sqrt{\frac{1-\omega}{M}}\big)$ as follows:
\begin{equation}\label{rewriteasfollow}
\begin{aligned}
C\Big(\sqrt{\frac{1-\omega}{M}}\Big)=&\sqrt[3]{-\mathcal{Q}+\sqrt{\mathcal{Q}^2+\mathcal{P}^3}}\\
&+\sqrt[3]{-\mathcal{Q}-\sqrt{\mathcal{Q}^2+\mathcal{P}^3}}+\mathcal{F},
\end{aligned}
\end{equation}
where $\mathcal{Q}=\frac{q}{2}$, $\mathcal{P}=\frac{p}{3}$ and $\mathcal{F}=-\frac{\beta}{3\alpha}$ by replacing $\mu$ as $\sqrt{\frac{1-\omega}{M}}$ in Theorem~\ref{theorem2}.

Meanwhile, the magnitudes of $\mathcal{Q}$ and $\mathcal{P}$ are mainly determined by their highest order terms, which can be verified by the following simulations. We denote the highest order terms in $\mathcal{Q}$ and $\mathcal{P}$ as $\hat{\mathcal{Q}}$ and $\hat{\mathcal{P}}$ respectively. The results are shown in Fig. \ref{aabbaabb}. When $M$ is large enough, the ratio $\frac{\mathcal{P}^3}{\mathcal{Q}^2}$ converges to $-1$, indicating that $\mathcal{Q}^2$ and $\mathcal{P}^3$ are on the same order and can be well approximated by their highest order terms $\hat{\mathcal{Q}}^2$ and $\hat{\mathcal{P}}^3$ respectively. Furthermore, when $M$ is large enough, $\hat{\mathcal{Q}}/\mathcal{Q}$ and $\hat{\mathcal{P}}/\mathcal{P}$ are equal to $1$, indicating that $\mathcal{Q}$ and $\mathcal{P}$ can be well approximated by their highest terms, i.e., $\hat{\mathcal{Q}}$ and $\hat{\mathcal{P}}$. In a word, we have
\begin{equation}\label{liminaword}
\left\{
\begin{aligned}
\lim_{M\rightarrow\infty}\mathcal{P}^3/\mathcal{Q}^2 & = \lim_{M\rightarrow\infty}\hat{\mathcal{P}}^3/\hat{\mathcal{Q}}^2 =-1 , \\
\lim_{M\rightarrow\infty}\hat{\mathcal{Q}}/\mathcal{Q} & =1 , \\
\lim_{M\rightarrow\infty}\hat{\mathcal{P}}/\mathcal{P} & =1 .
\end{aligned}
\right.
\end{equation}

Based on the above arguments, we obtain
\begin{align}
&\lim_{M\rightarrow\infty}C\bigg(\sqrt{\frac{1-\omega}{M}}\bigg)\nonumber\\
=&\sqrt[3]{-\frac{1}{729}\Big(\frac{M}{1-\omega}\Big)^{\frac{3}{2}}+\sqrt{\Big(\big(\frac{1}{729}\big)^2+\big(-\frac{1}{81}\big)^3\Big)\Big(\frac{M}{1-\omega}\Big)^3}}\nonumber\\
+&\sqrt[3]{-\frac{1}{729}\Big(\frac{M}{1-\omega}\Big)^{\frac{3}{2}}-\sqrt{\Big(\big(\frac{1}{729}\big)^2+\big(-\frac{1}{81}\big)^3\Big)\Big(\frac{M}{1-\omega}\Big)^3}}\nonumber\\
+&\frac{8}{9}\sqrt{\Big(\frac{M}{1-\omega}\Big)}+\frac{2}{3}=\frac{2}{3}\bigg(\sqrt{\frac{M}{1-\omega}}+1\bigg).\label{C1DELABEL}
\end{align}

\begin{figure}[htbp]
\centering
\subfigure[Results of $\hat{\mathcal{Q}}/\mathcal{Q}$ and $\hat{\mathcal{P}}/\mathcal{P}$.]{
\begin{minipage}[t]{0.5\linewidth}
\centering
\includegraphics[width=1.75in]{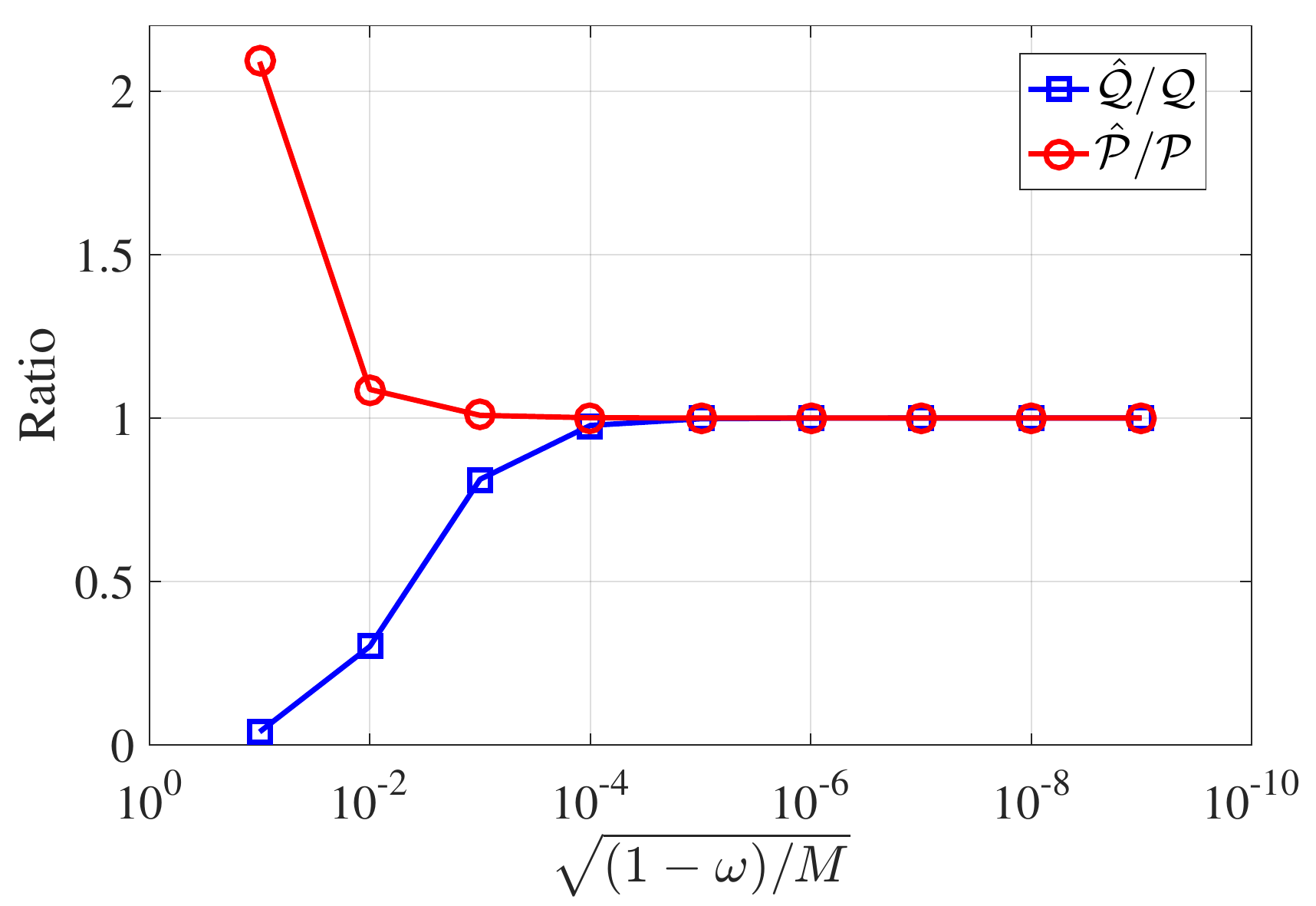}
\end{minipage}%
}%
\subfigure[Results of $\mathcal{P}^3/\mathcal{Q}^2$ and $\hat{\mathcal{P}}^3/\hat{\mathcal{Q}}^2$.]{
\begin{minipage}[t]{0.5\linewidth}
\centering
\includegraphics[width=1.8in]{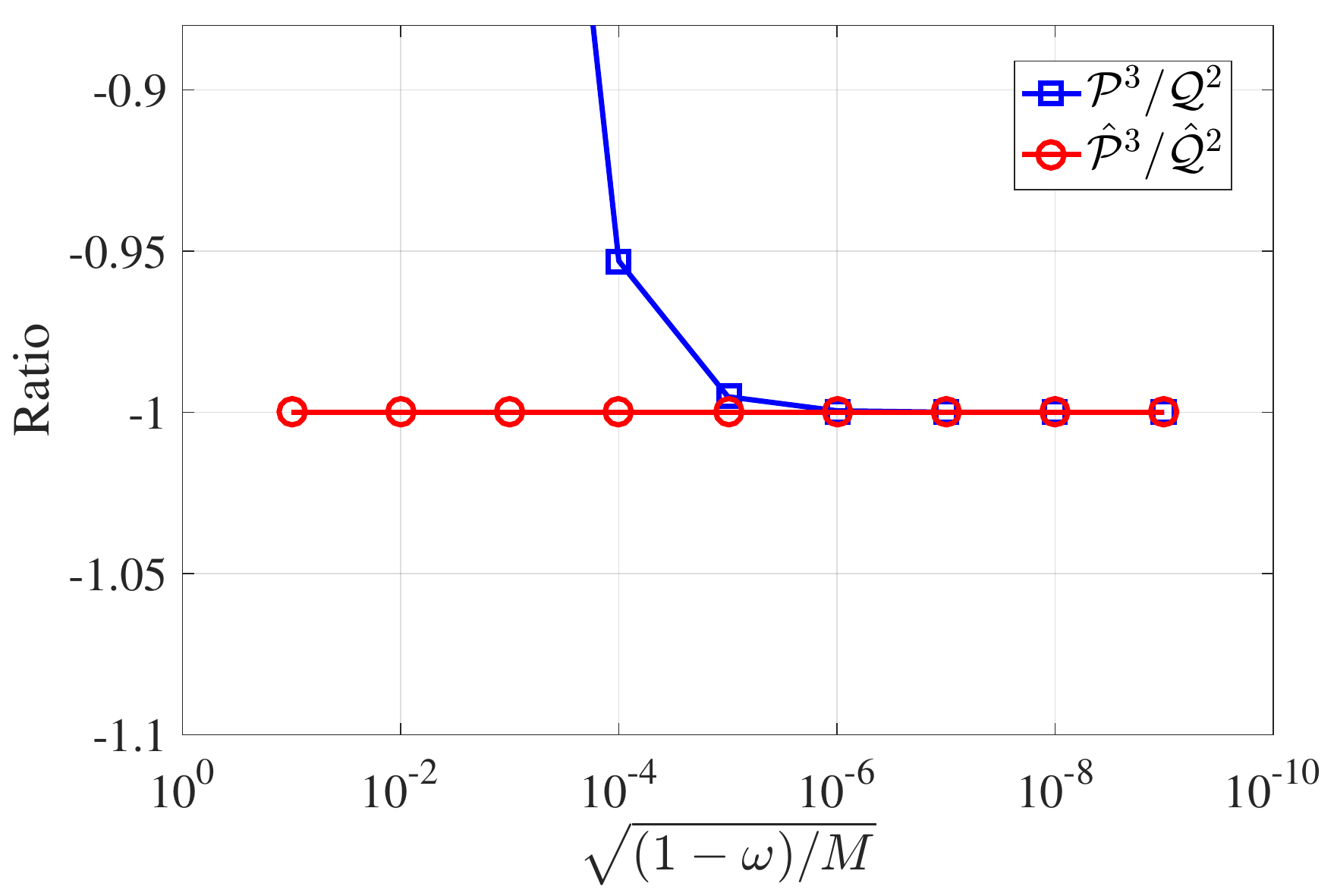}
\end{minipage}%
}%
\centering
\caption{Ratios of different parameters with the decrease of $\sqrt{(1-\omega)/M}$.}\label{aabbaabb}
\end{figure}

\end{IEEEproof}

\begin{rmk5}
Similar to the derivation in Remark \ref{rmk}, we have
\begin{equation}\label{theintuitivecondition}
\lim_{\mu\rightarrow0} C(\mu)=\frac{2}{3}\Big(\frac{1}{\mu}+1\Big).
\end{equation}
It is known that the MIP-based sufficient condition for the ERC of OMP derived by Tropp \cite{2} is
\begin{equation}\label{Tropp}
K<\frac{1}{2}\Big(\frac{1}{\mu}+1\Big).
\end{equation}
Therefore, the sufficient condition for the ERC of OLS and MOLS based on (\ref{theintuitivecondition}), i.e.,
\begin{equation}\label{comtropp}
K<\frac{2}{3}\Big(\frac{1}{\mu}+1\Big)
\end{equation}
improves the result (\ref{Tropp}) in the case of low coherence, which indicates OLS and MOLS algorithms can reconstruct more exactly than OMP algorithm from the perspective of
reconstructible sparsity level.
\label{rmk5}
\end{rmk5}

It is proved in \cite{89} that the optimal value of MIP-based sufficient condition for decaying signals is $K<\frac{1}{\mu}$. Our result in Remark \ref{rmk5} indicates that $K<\frac{2}{3}(\frac{1}{\mu}+1)$ is sufficient for general sparse signals, which means that (\ref{comtropp}) is at most $\frac{2}{3}$ times worse than the optimal value. However, our analysis applies for general sparse signals rather than for only decaying sparse signals. Moreover, Remark \ref{rmk5} shows that our derived result improves that in \cite{2}. That is to say, our result exhibits stronger generality and superiority.

\subsection{ERCs for BOLS algorithm exploiting block structure}

The main difference of the analysis between BOLS and the aforementioned two algorithms is that we consider the block structure within the sparse signals. To do so, we adopt two independent concepts of coherence: block-coherence that indicates the inter-block global coherence properties of the measurement matrix $\mathbf{D}$, and sub-coherence that captures the intra-block local coherence properties of $\mathbf{D}$.

\begin{def3}
(Block-coherence \cite{1})
The block-coherence of $\mathbf{D}$ is defined as
\begin{equation}\label{block coherence}
\mu_B=\max_{i,j\neq i}\frac{\rho(\mathbf{M}[i,j])}{d},
\end{equation}
where $\mathbf{M}[i,j]=\mathbf{D}^T[i]\mathbf{D}[j]$. It is similar to (\ref{coherence}) by replacing columns $\mathbf{D}_i$ of $\mathbf{D}$ by its sub-blocks $\mathbf{D}[i]$.
\label{definitionofblockcoherence}
\end{def3}

\begin{def4}
(Sub-coherence \cite{1})
The sub-coherence of $\mathbf{D}$ is defined as
\begin{equation}\label{sub block coherence}
\nu=\max_l\max_{i,j\neq i}|<\mathbf{D}_i,\mathbf{D}_j>|,\quad\mathbf{D}_i,\mathbf{D}_j\in \mathbf{D}[l].
\end{equation}
\label{definitionofsubcoherence}
\end{def4}

Based on aforementioned definitions, we derive the following Lemmas.

\begin{lemma2}
Let $k$ be the block sparsity level, $d$ be the block length, $\mu$ be the matrix coherence and $\nu$ be the sub-coherence. Suppose $\mu<\frac{1}{kd-1}$, then $\frac{1}{\sqrt{\mathcal{T}_B}}\leq||\mathbf{P}^\bot_{\mathbf{S}^{l}}\mathbf{D}_i||_2\leq1$ for $i\in\{1,2,\cdots,N\}\backslash \mathbf{S}^l$,
where $\mathcal{T}_B=\Big(1-\Big(\frac{\sqrt{1+(kd-1)\mu}\sqrt{kd\mu^2}}{1-(d-1)\nu-(k-1)d\mu}\Big)^2\Big)^{-1}$.
\label{lemma2}
\end{lemma2}

\begin{IEEEproof}
See Appendix \ref{proofoflemma2}.
\end{IEEEproof}
\begin{lemma5}
For a random matrix $\mathbf{B}\in \mathcal{R}^{M\times kd}$ $(k\geq2)$ whose entries are independently and identically distributed as $\mathcal{N}(0,\frac{1}{M})$, the following probability holds:
\begin{equation}\label{probability}
{\rm P}\Big\{\rho_c(\mathbf{B}^T\mathbf{B}-\mathbf{I})\leq \frac{(k-\mathcal{T}_B)d\tau}{2}\Big\}\geq1-\frac{ke^{-\frac{kd-1}{2}(\mathcal{C}-\log(1+\mathcal{C}))}}{\mathcal{C}\sqrt{\pi(kd-1)}},
\end{equation}
where $\tau$ is a positive constant, $d$ is the block length and $\mathcal{C}=\frac{M(k-\mathcal{T}_B)^2d^2\tau^2}{4(kd-1)^2}-1$.
\label{lemma5}
\end{lemma5}

The proof is omitted since it is similar to that of Lemma \ref{lemma4}.

\begin{rmk7}
Similar to Remark \ref{rmk6}, we have the following probability:
\begin{equation}\label{probabilitybols}
\lim_{M\rightarrow\infty} 1-\frac{ke^{-\frac{kd-1}{2}(\mathcal{C}-\log(1+\mathcal{C}))}}{\mathcal{C}\sqrt{\pi(kd-1)}}=1.
\end{equation}
\label{rmk7}
\end{rmk7}

Intuitively, we give a geometric interpretation of BOLS in terms of orthogonal projections as shown in Proposition \ref{proposition1} below to facilitate the following derivation.
\begin{proposition1}
In the $(l+1)$-th iteration, the BOLS algorithm identifies a block support index:
\begin{align}
i_B^{l+1}&=\mathop{\arg\max}\limits_{j_B\in\{1,..,N_B\}\backslash\mathbf{S}_B^{l}}\Bigg(\frac{|<\mathbf{D}_{(j_B-1)d+1},\mathbf{r}^{l}>|}{||\mathbf{P}^\bot_{\mathbf{S}^{l}}\mathbf{D}_{(j_B-1)d+1}||_2}\Bigg)^2\nonumber\\
&+\sum^{j_Bd}_{j=(j_B-1)d+2}\Bigg(\frac{|<\mathbf{D}_j,\mathbf{r}^{l}>|}{||\mathbf{P}^\bot_{\mathbf{S}^{l}\cup\{(j_B-1)d+1,\cdots,j-1\}}\mathbf{D}_j||_2}\Bigg)^2.\label{ctri}
\end{align}
\label{proposition1}
\end{proposition1}
\begin{IEEEproof}
See Appendix \ref{proofofproposition1}.
\end{IEEEproof}

The ERC and the MIP-based sufficient condition for BOLS are given in Theorem \ref{theorem5} and Theorem \ref{theorem6}.

\begin{theorem5}
Assume that BOLS has chosen $l$ correct blocks after $l$ iterations. A sufficient condition for BOLS to select one correct block in the $(l+1)$-th iteration is
\begin{equation}\label{sufficient condition}
\rho_c((\mathbf{Q}_{0\backslash \mathbf{S}}\mathbf{R}_{0\backslash \mathbf{S}})^\dag(\overline{\mathbf{Q}}_0\overline{\mathbf{R}}_0))<1,
\end{equation}
where $\mathbf{R}_{0\backslash\mathbf{S}}$ and $\overline{\mathbf{R}}_0$ are given by
\begin{equation}\label{matrix1}
\begin{aligned}
\mathbf{R}_{0\backslash \mathbf{S}}=&{\rm diag}\Big(\frac{1}{||\mathbf{P}^\bot_{\mathbf{S}^{l}}\mathbf{D}_1||_2},\frac{1}{||\mathbf{P}^\bot_{\mathbf{S}^{l}\cup\{1\}}\mathbf{D}_2||_2},\cdots,\\
&\frac{1}{||\mathbf{P}^\bot_{\mathbf{S}^{l}\cup\{1,\cdots,d-1\}}\mathbf{D}_{d}||_2},\cdots\Big)
\end{aligned}
\end{equation}
and
\begin{align}
\overline{\mathbf{R}}_0=&{\rm diag}\Big(\frac{1}{||\mathbf{P}^\bot_{\mathbf{S}^{l}}\mathbf{D}_{kd+1}||_2},\frac{1}{||\mathbf{P}^\bot_{\mathbf{S}^{l}\cup\{kd+1\}}\mathbf{D}_{kd+2}||_2},\cdots,\nonumber\\
&\frac{1}{||\mathbf{P}^\bot_{\mathbf{S}^{l}\cup\{(2k-1)d+1,\cdots,2kd-1\}}\mathbf{D}_{2kd}||_2},\cdots\Big),\label{Rofeiblock}
\end{align}
if $\{1,\cdots,kd\}$ are support indices in $\mathbf{Q}_0$ and $\{1,\cdots,d\}\not\subset \mathbf{S}^l$.
\label{theorem5}
\end{theorem5}
\begin{IEEEproof}
See Appendix \ref{proofoftheorem5}.
\end{IEEEproof}

%
%

%
%

\begin{theorem6}
Let $\mu$ be the matrix coherence, $\mu_B$ be the block-coherence and $\nu$ be the sub-coherence of the measurement matrix $\mathbf{D}$. The ERC (\ref{sufficient condition}) is satisfied if
\begin{equation}\label{theorem6mian}
kd<\Big(\sqrt[3]{-\frac{q_B}{2}+\sqrt{\Delta_B}}+\sqrt[3]{-\frac{q_B}{2}-\sqrt{\Delta_B}}-\frac{\beta_B}{3\alpha_B}\Big),
\end{equation}
where $q_B=\frac{27\alpha_B^2\delta_B-9\alpha_B\beta_B\gamma_B+2\beta_B^3}{27\alpha_B^3}$, $p_B=\frac{3\alpha_B\gamma_B-\beta_B^2}{3\alpha_B^2}$, $\alpha_B=-\mu_B\mu^3+3\mu_B\mu^2$, $\beta_B=(2+\mu_B)\mu^3-(d\mu_B+\mu_B+2)\mu^2+6\mu_B\mathcal{G}\mu$, $\gamma_B=-2\mu^3+2\mu^2-(2d\mu_B\mathcal{G}+4\mathcal{G})\mu+3\mu_B\mathcal{G}^2$, $\delta_B=-d\mu_B\mathcal{G}^2-2\mathcal{G}^2$, $\mathcal{G}=(d-1)\nu-1-d\mu$ and $\Delta_B=(\frac{q_B}{2})^2+(\frac{p_B}{3})^3$.
\label{theorem6}
\end{theorem6}

\begin{IEEEproof}
See Appendix \ref{proofoftheorem6}.
\end{IEEEproof}

\begin{rmk3}
Theorem \ref{theorem6} applies to OLS when $d=1$, which is identical to Theorem~\ref{theorem2}.
\label{rmk3}
\end{rmk3}

While the main result in Theorem \ref{theorem6} appears to be cumbersome, it accurately quantifies the performance bounds of BOLS and offers adequate intuitions for a given problem setting. Along this line, Remark \ref{rmk4} arises.

\begin{rmk4}
To clearly observe the relationship between the upper bound of sparsity and matrix coherence, and the relationship between the upper bound of sparsity and block-coherence, let us consider the special case of a block structure matrix, that is, the sub-coherence satisfies that $\nu=0$. Denote the right-side of (\ref{theorem6mian}) as $C_B(\mu,\mu_B)$. The upper bound of $C_B(\mu,\mu_B)$ is $C_B\big(\sqrt{\frac{N-M}{M(N-1)}},\frac{1}{d}\sqrt{\frac{N-M}{M(N-1)}}\big)$. Depending on the assumptions adopted, the corresponding analytical results for the upper bound are given as follows. (1) When $M$ is fixed, the upper bound of $C_B(\mu,\mu_B)$ decreases with
the increase of $N$. (2) When $N$ is fixed, the upper bound of
$C_B(\mu,\mu_B)$ is improved with the increase of $M$.
(3) Letting $\omega=\frac{M}{N}$, and for $M\gg0$ and $N\gg0$,
the upper bound of the theoretical threshold $C_B(\mu,\mu_B)$ satisfies
\begin{equation}\label{cbupperbound}
\begin{aligned}
C_B(\mu,\mu_B)\leq \mathcal{W}(d)\sqrt{\frac{M}{1-\omega}}+\frac{5}{9}d+\frac{1}{9},
\end{aligned}
\end{equation}
where
\begin{equation}\label{wd=}
\begin{aligned}
\mathcal{W}(d)=&\sqrt[3]{-\mathcal{U}+\sqrt{\mathcal{U}^2+\mathcal{V}^3}}\\
&+\sqrt[3]{-\mathcal{U}-\sqrt{\mathcal{U}^2+\mathcal{V}^3}}+\frac{2}{9}d+\frac{2}{3},
\end{aligned}
\end{equation}
$\mathcal{U}=-\frac{8}{729}d^3+\frac{4}{81}d^2-\frac{2}{27}d+\frac{1}{27}$, and $\mathcal{V}=-\frac{4}{81}d^2+\frac{4}{27}d-\frac{1}{9}$.

\label{rmk4}
\end{rmk4}

\begin{IEEEproof}
By direct calculations, the theoretical threshold $C_B(\mu,\mu_B)$ is improved with the decrease of $\mu$ or $\mu_B$, i.e., $\nabla C_B(\mu)<0$ and $\nabla C_B(\mu_B)<0$. In order to obtain the upper bound of $C_B(\mu,\mu_B)$, it is necessary to derive the minimum value of $\mu_B$.

For a given matrix $\mathbf{Q}\in \mathcal{R}^{d\times d}$,
\begin{equation}\label{roahahah}
\rho(\mathbf{Q})\geq||\mathbf{Q}||_{\max},
\end{equation}
where $||\cdot||_{\max}$ represents the largest element of its objective.
Therefore, the block-coherence of a given matrix $\mathbf{D}$ satisfies
\begin{equation}\label{blockcoherencerelation}
\mu_B\geq\frac{\mu}{d}\geq\frac{1}{d}\sqrt{\frac{N-M}{M(N-1)}},
\end{equation}

Finally, similar to the proof of Remark \ref{rmk}, Remark \ref{rmk4} is concluded.
\end{IEEEproof}

In general, the derived upper bound is improved with the increase of $d$ and $M$. On the one hand, the results mean that the performance of the algorithm generally increases with the increase of the number of measurements. On the other hand, the larger $d$ makes the block structure more powerful, leading to better performance of BOLS.
\begin{rmk8}
Define the right-side of (\ref{theorem6mian}) as $C_B(\mu,\mu_B,\nu)$. In asymptotic case, we have
\begin{equation}\label{BOLSasymptotic}
\lim\limits_{\mu_B,\nu\rightarrow0;\;\mu\rightarrow \mu_B}C_B(\mu,\mu_B,\nu)=\frac{2}{3}\Big(\frac{1}{\mu_B}+\frac{7}{6}d-\frac{1}{6}\Big).
\end{equation}

The well-known MIP-based sufficient condition for the ERC of BOMP by using orthogonal block uncertainty relations in \cite{1} is
\begin{equation}\label{well-knownBOMP}
kd<\frac{1}{2}\Big(\frac{1}{\mu_B}+d\Big).
\end{equation}
Since $d\geq1$ and $\mu_B\in[0,1]$, the sufficient condition based on (\ref{BOLSasymptotic}), i.e.,
\begin{equation}\label{bolsimprov}
kd<\frac{2}{3}\Big(\frac{1}{\mu_B}+\frac{7}{6}d-\frac{1}{6}\Big),
\end{equation}
is improved compared with (\ref{well-knownBOMP}). This result implies that BOLS performs better than BOMP from the perspective of reconstructible sparsity level.
\label{rmk8}
\end{rmk8}



\section{Noisy Recovery Conditions}
\label{section4}

In the noisy case, to guarantee reliable recovery, the SNR should be sufficiently high for correct atom selection. Meanwhile, the level of MIP needs to be small, otherwise the atoms cannot be separated well.
Considering these issues, this section presents the recovery conditions in the noisy case for the OLS-type algorithms based on the theoretical results for the noiseless case in Theorem \ref{theorem2} and Theorem \ref{theorem6}.

The noisy form of the CS model (\ref{CSmodel}) is given by
\begin{equation}\label{systemnoisy}
\mathbf{y}=\mathbf{D}\mathbf{x}+\mathbf{\epsilon},
\end{equation}
where $\mathbf{\epsilon}\sim\mathcal{N}(0,\sigma^2\mathbf{I}_M)$ is the Gaussian noise.
Note that the residual in the $l$-th iteration is
\begin{equation}\label{residual}
\mathbf{r}^{l}=(\mathbf{I}-\mathbf{P}_{\mathbf{S}^l})\mathbf{y}=(\mathbf{I}-\mathbf{P}_{\mathbf{S}^l})\mathbf{D}\mathbf{x}+(\mathbf{I}-\mathbf{P}_{\mathbf{S}^l})\mathbf{\epsilon}.
\end{equation}
Define $\mathbf{s}^l=(\mathbf{I}-\mathbf{P}_{\mathbf{S}^l})\mathbf{D}\mathbf{x}$ and $\mathbf{n}^l=(\mathbf{I}-\mathbf{P}_{\mathbf{S}^l})\mathbf{\epsilon}$ as the signal and noise parts of the residual, respectively. Then the analysis of recovery conditions for OLS-type algorithms under the noisy model (\ref{systemnoisy}) is investigated in the following subsections.

\subsection{Noisy recovery conditions for OLS and MOLS algorithms}

To obtain the recovery conditions for the OLS and MOLS algorithms in the noisy case, we first present the following lemma.
\begin{lemma6}
If the noiseless condition in (\ref{theorem2mian2}) holds, then
\begin{equation}\label{lemma6main}
||(\mathbf{Q}_{0\backslash\mathbf{S}}\mathbf{R}_{0\backslash\mathbf{S}})^\dag(\overline{\mathbf{Q}}_{0}\overline{\mathbf{R}}_{0})||_{1,1}\leq\frac{2K\mathcal{T}\mu}{2-(K-\mathcal{T})\mu}<1.
\end{equation}
\label{lemma6}
\end{lemma6}
The proof of Lemma \ref{lemma6} follows directly from the proof of Theorem \ref{theorem2} in Appendix \ref{proofoftheorem2}. Then, based on Lemma \ref{lemma6}, the following theorem holds.

\begin{theorem7}
Suppose that the condition in (\ref{theorem2mian2}) holds and the remaining nonzero vector $\mathbf{x}_{0\backslash\mathbf{S}}$ in the $(l+1)$-th iteration satisfies
\begin{equation}\label{remainingentries}
||\mathbf{x}_{0\backslash\mathbf{S}}||_{2}>\frac{2\sqrt{K-l}(2-(K-\mathcal{T})\mu)\sigma\sqrt{M+2\sqrt{M\log M}}}{(2-(K-\mathcal{T})\mu-2K\mathcal{T}\mu)(1-(K-1)\mu)}.
\end{equation}
Then, in the noisy case, the OLS algorithm chooses one correct atom and the MOLS algorithm chooses at least one correct atom in the $(l+1)$-th iteration with probability at least $1-\frac{1}{M}$.
\label{theorem7}
\end{theorem7}
\begin{IEEEproof}
See Appendix I.
\end{IEEEproof}

Based on Theorem \ref{theorem7}, we have the subsequent corollary.
\begin{Corollary1}
Suppose that the condition in (\ref{theorem2mian2}) holds and all the nonzero entries $\mathbf{x}_i$ satisfy
\begin{equation}\label{remainingentries}
|\mathbf{x}_i|>\frac{2(2-(K-\mathcal{T})\mu)\sigma\sqrt{M+2\sqrt{M\log M}}}{(2-(K-\mathcal{T})\mu-2K\mathcal{T}\mu)(1-(K-1)\mu)}.
\end{equation}
Then, in the noisy case, the OLS and MOLS algorithms choose the true support set with probability at least $1-\frac{1}{M}$.
\label{corollary1}
\end{Corollary1}

Theorem \ref{theorem7} and Corollary \ref{corollary1} indicate that if the nonzero entries are large enough, the OLS and MOLS algorithms choose the correct atoms in the noisy case.

\subsection{Noisy recovery conditions for BOLS algorithm}
Similarly, to derive the recovery conditions for the BOLS algorithm in the noisy case, we give the following lemmas.
\begin{lemma7}
If the noiseless condition in (\ref{theorem6mian}) holds, then
\begin{equation}\label{lemma6main}
\rho_c((\mathbf{Q}_{0\backslash \mathbf{S}}\mathbf{R}_{0\backslash \mathbf{S}})^\dag(\overline{\mathbf{Q}}_{0}\overline{\mathbf{R}}_{0}))<\frac{2\mathcal{T}_Bkd\mu_B}{2-(k-\mathcal{T}_B)d\mu_B}<1.
\end{equation}
\label{lemma7}
\end{lemma7}
\begin{lemma8}
The Gaussian noise $\mathbf{\epsilon}\sim\mathcal{N}(0,\sigma^2\mathbf{I}_M)$ satisfies
\begin{equation}\label{gaussiannoise}
{\rm P}\Big\{||\mathbf{D}^T\mathbf{\epsilon}||_{2,\infty}\leq\sqrt{d}\sigma\sqrt{M+2\sqrt{M\log M}}\Big\}\geq1-\frac{1}{M},
\end{equation}
where $d$ is the block length.
\label{lemma8}
\end{lemma8}
The proof of Lemma~\ref{lemma8} can be extended from that of Lemma~5.1 in \cite{86}.
Based on Lemmas~\ref{lemma7} and~\ref{lemma8}, we obtain the following Theorem~\ref{theorem8}.
\begin{theorem8}
Suppose that the condition in (\ref{theorem6mian}) holds and the remaining nonzero vector $\mathbf{x}_{0\backslash\mathbf{S}}$ in the $(l+1)$-th iteration satisfies
\begin{equation}\label{remainingentries}
\begin{aligned}
&||\mathbf{x}_{0\backslash\mathbf{S}}||_{2}\\
>&\frac{2\sqrt{k-l}(2-(k-\mathcal{T}_B)d\mu_B)\sqrt{d}\sigma\sqrt{M+2\sqrt{M\log M}}}{(2-(k-\mathcal{T}_B)d\mu_B-2\mathcal{T}_Bkd\mu_B)(1-(kd-1)\mu)}.
\end{aligned}
\end{equation}
Then, in the noisy case, the BOLS algorithm chooses one correct block in the $(l+1)$-th iteration with probability at least $1-\frac{1}{M}$.
\label{theorem8}
\end{theorem8}

\begin{IEEEproof}
See Appendix J.
\end{IEEEproof}

The following corollary is derived from Theorem~\ref{theorem8}.
\begin{Corollary2}
Suppose that the condition in (\ref{theorem6mian}) holds and all the nonzero blocks $\mathbf{x}[i]$ satisfy
\begin{equation}\label{remainingentriesbols}
\begin{aligned}
||\mathbf{x}[i]||_2>\frac{2(2-(k-\mathcal{T}_B)d\mu_B)\sqrt{d}\sigma\sqrt{M+2\sqrt{M\log M}}}{(2-(k-\mathcal{T}_B)d\mu_B-2\mathcal{T}_Bkd\mu_B)(1-(kd-1)\mu)}.
\end{aligned}
\end{equation}
Then, in the noisy case, the BOLS algorithm chooses the true support set with probability at least $1-\frac{1}{M}$.
\label{corollary2}
\end{Corollary2}

\section{Simulation Tests}
In this section, we present simulation tests to illustrate our theoretical results shown in Section \ref{section3} and Section IV, and compare them with the existing ones.

\subsection{Simulation tests for noiseless recovery conditions}

\subsubsection{Comparisons of the lower bounds of $||\mathbf{P}^{\bot}_{\mathbf{S}^l}\mathbf{D}_i||_2$}
In this subsection, we exploit simulations to illustrate Lemma~\ref{lemma3}, Lemma \ref{lemma2} and compare them with (\ref{lemma2and5}) \cite{85}.

\begin{figure}[htbp]
\centering
\subfigure[The bounds versus $\mu$.]{
\begin{minipage}[t]{0.5\linewidth}
\centering
\includegraphics[width=1.8in]{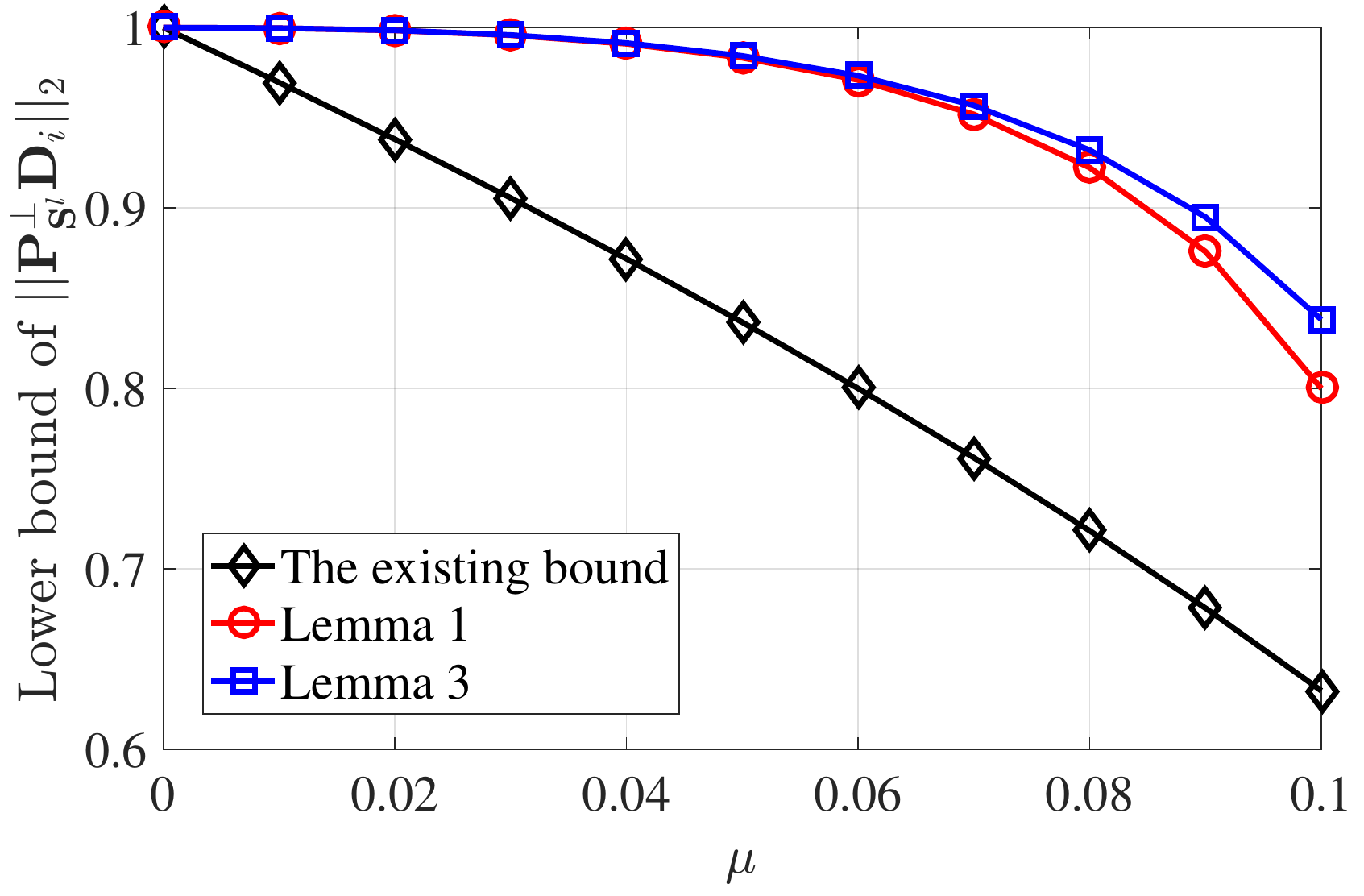}
\end{minipage}%
}%
\subfigure[The bounds versus $k$.]{
\begin{minipage}[t]{0.5\linewidth}
\centering
\includegraphics[width=1.8in]{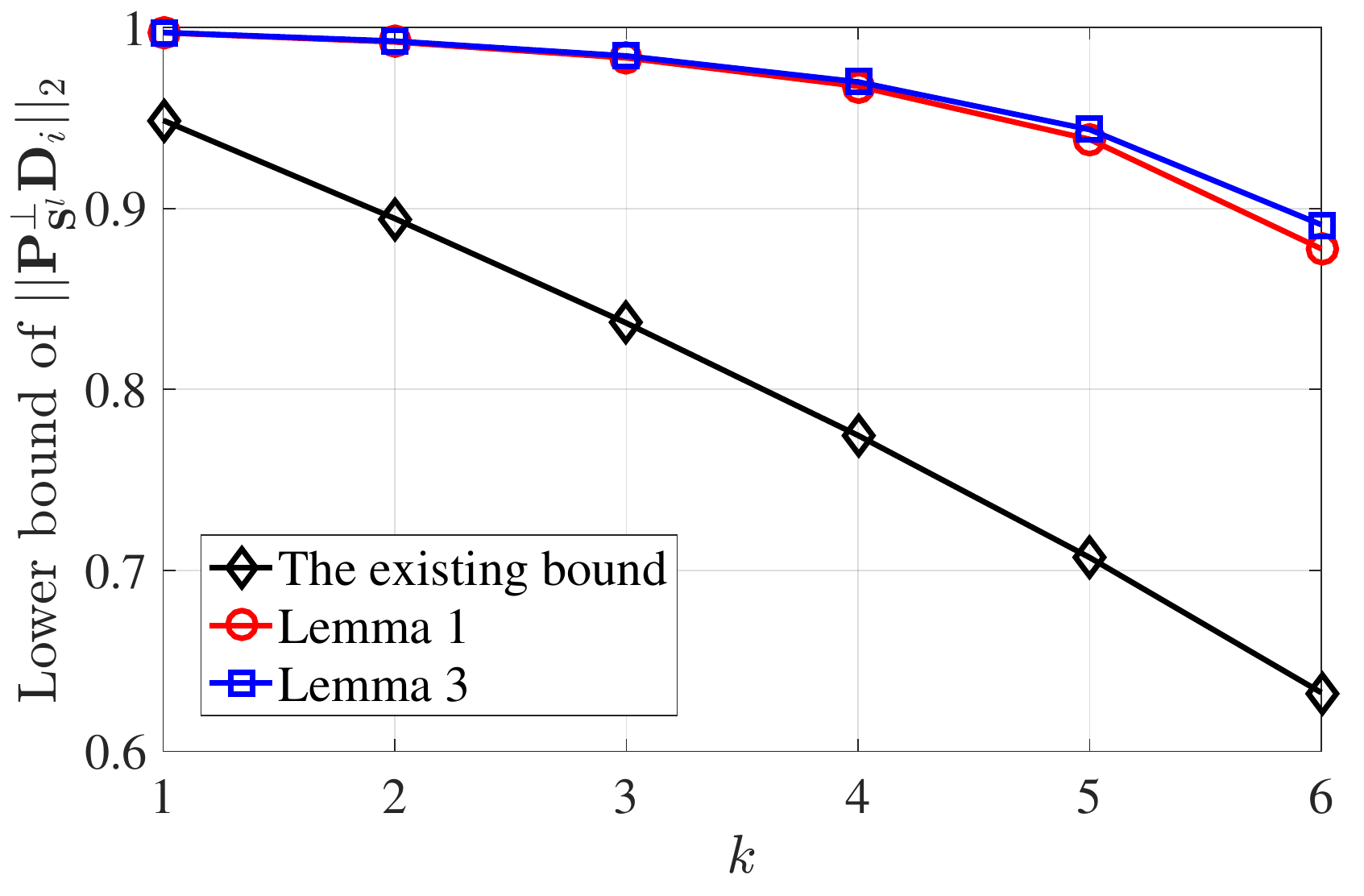}
\end{minipage}%
}%
\centering
\caption{Comparisons of lower bounds of $||\mathbf{P}^{\bot}_{\mathbf{S}^l}\mathbf{D}_i||_2$ with (a) $K=6$, $k=3$, $d=2$ and $\nu=\frac{1}{2}\mu$; (b) $K=kd$, $d=2$, $\mu=0.05$ and $\nu=\frac{1}{2}\mu$.}\label{lowerbounds}
\end{figure}

The simulation parameters satisfy $\mu(K-1)<1$ and $\mu(kd-1)<1$, which are given in Lemma~\ref{lemma3}, Lemma~\ref{lemma2} and the precondition of (\ref{lemma2and5}).
Fig. \ref{lowerbounds} shows that our derived results are much tighter than the existing bound which indicates that the follow-up theoretical results based on these bounds, such as Theorems~\ref{theorem2} and \ref{theorem6}, are much sharper than the one using (\ref{lemma2and5}). Meanwhile, it is observed that the lower bounds using sub-coherence in Lemma~\ref{lemma3} is much tighter than that just using conventional matrix coherence. This means the characteristics of block structure plays a role in improving the boundary.

\subsubsection{Comparisons of the sufficient conditions for ERCs}
The simulations of Theorem \ref{theorem2} and Theorem \ref{theorem6} are conducted. We compare Theorem \ref{theorem2} with
Theorem 3.5 in \cite{2}, and Theorem~\ref{theorem6} with Theorem~3 in \cite{1} respectively.

\begin{figure}
  \centering
  \includegraphics[scale=0.34]{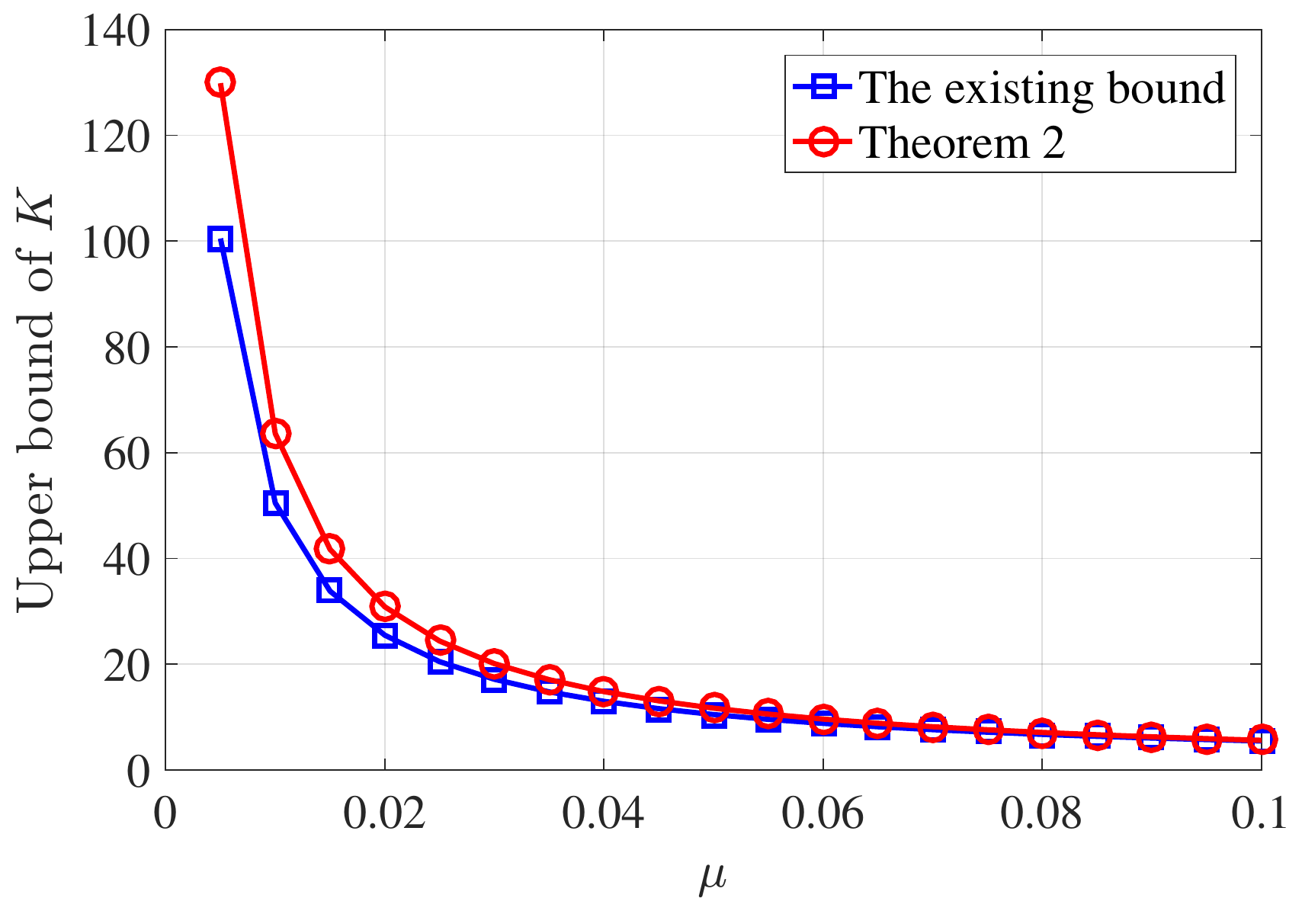}\\
  \caption{Comparisons of upper bounds of $K$ as a function of $\mu$.}\label{OLSUPPER}
\end{figure}

\begin{figure}
  \centering
  \includegraphics[scale=0.34]{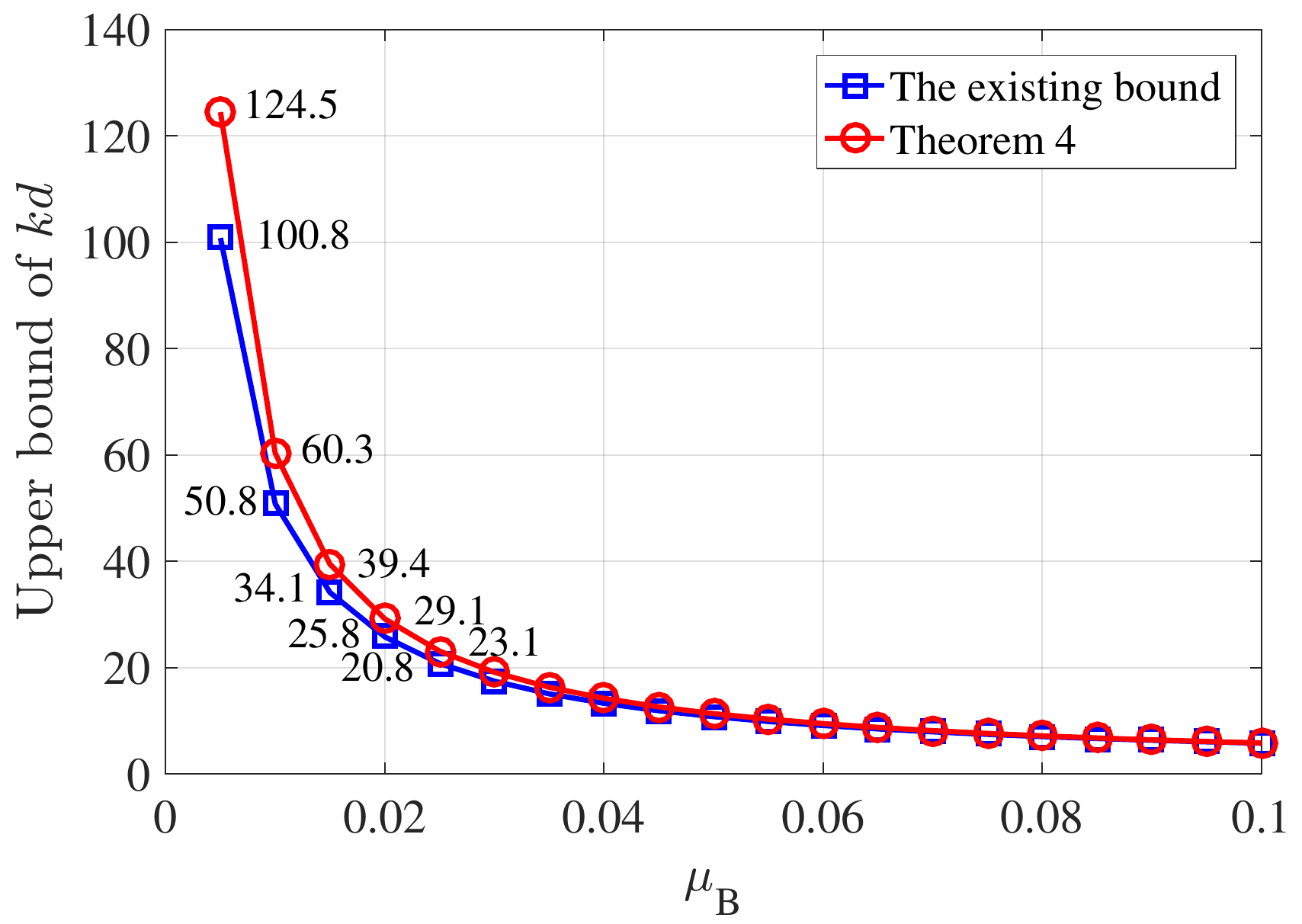}\\
  \caption{Comparisons of upper bounds of $kd$ as a function of $\mu_B$ with $\mu=1.2\mu_B$, $\nu=0.8\mu_B$ and $d=4$.}\label{BOLSUPPERD=4}
\end{figure}

\begin{figure}
  \centering
  \includegraphics[scale=0.34]{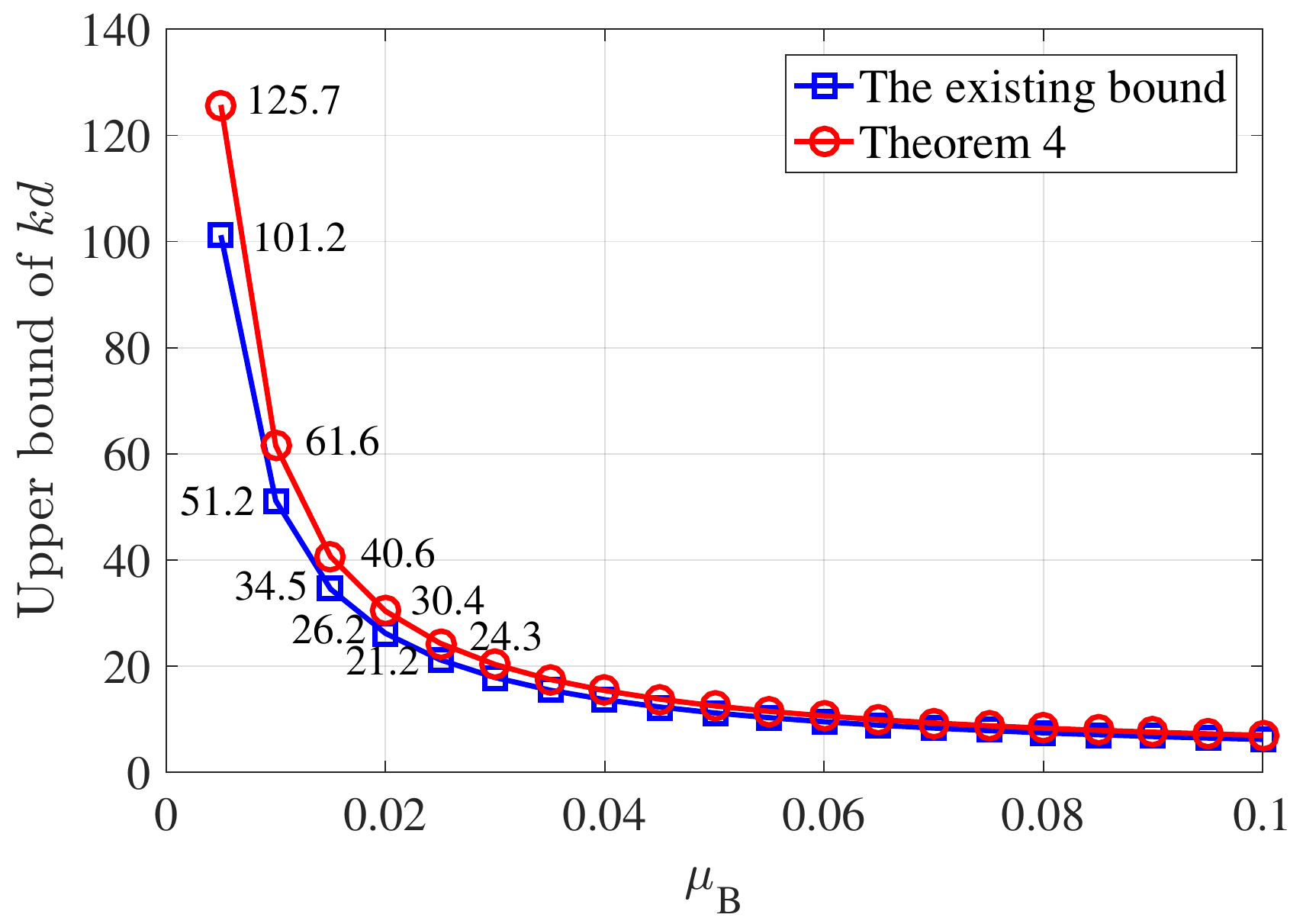}\\
  \caption{Comparisons of upper bounds of $kd$ as a function of $\mu_B$ with $\mu=1.2\mu_B$, $\nu=0.8\mu_B$ and $d=8$.}\label{BOLSUPPERD=8}
\end{figure}

Figs. \ref{OLSUPPER}-\ref{BOLSUPPERD=8} respectively present the existing bounds and our derived bounds for OLS, MOLS, BOLS with $d=4$ and BOLS with $d=8$. The results show that the upper bounds of $K$ and $kd$ in Theorem \ref{theorem2} and Theorem \ref{theorem6} are higher than the existing bounds under the same parameter settings, resulting in more reliable reconstructible sparsity levels of OLS-type algorithms than those of OMP-type algorithms. They also show that for the block sparse signals, the upper bounds of $kd$ are improved with the increase of $d$.

\subsection{Simulation tests for noisy recovery conditions}
In this subsection, we perform simulations to illustrate Theorem \ref{theorem7}, Theorem \ref{theorem8}, and compare them with the existing bound (Eqn. 10 \cite{85}).

\begin{figure}[htbp]
\centering
\subfigure[]{
\begin{minipage}[t]{0.5\linewidth}
\centering
\includegraphics[width=1.8in]{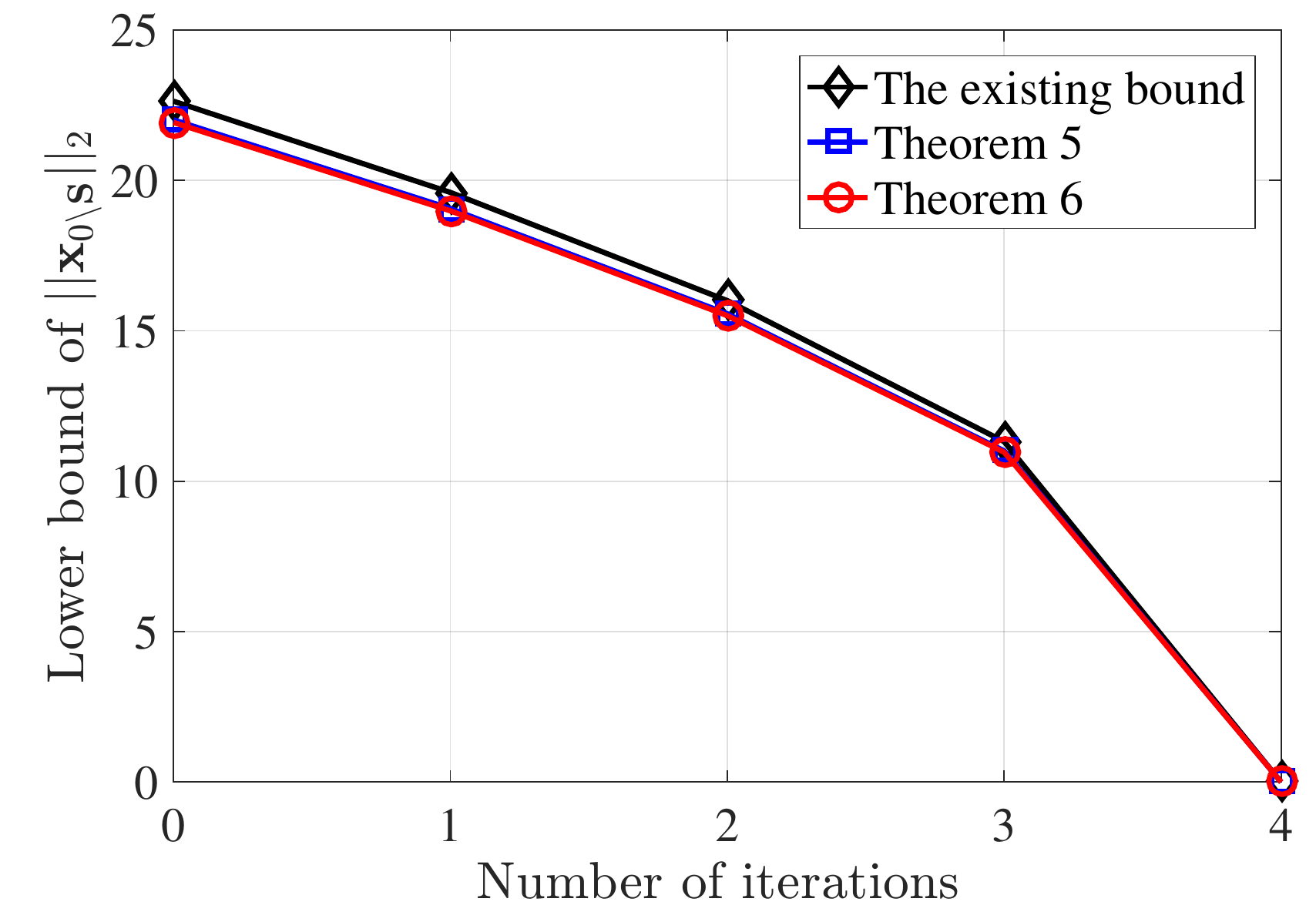}
\end{minipage}%
}%
\subfigure[]{
\begin{minipage}[t]{0.5\linewidth}
\centering
\includegraphics[width=1.8in]{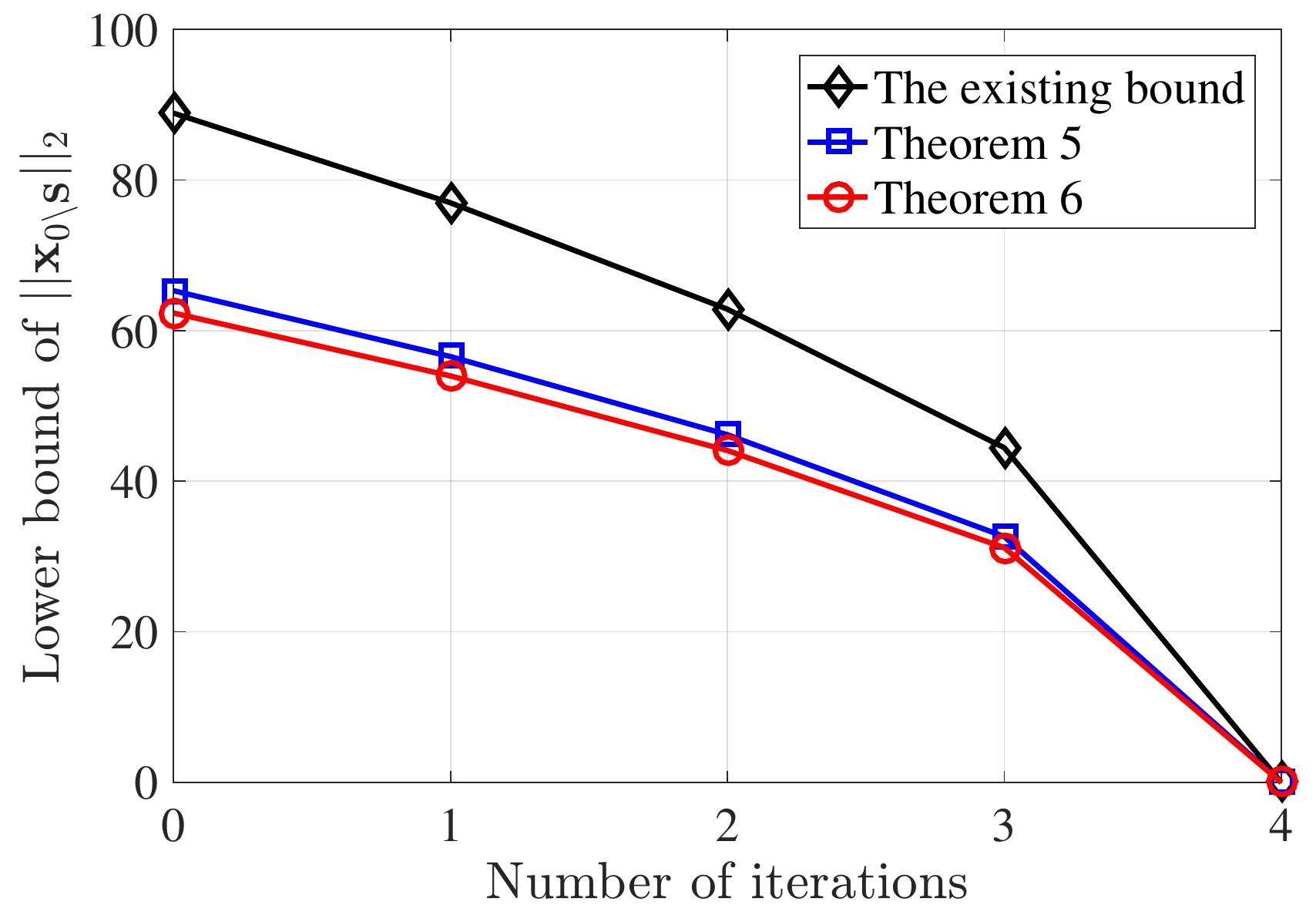}
\end{minipage}%
}%
\centering
\caption{Comparisons of lower bounds of $||\mathbf{x}_{0\backslash\mathbf{S}}||_{2}$ as a function of $l$ with $\mu_B=0.025$, $\nu=\mu_B$, $k=4$, $\sigma=0.1$, $M=512$ and (a) $d=2$; (b) $d=4$.}\label{lowerboundsx}
\end{figure}

For fair comparison, $\mu$ and $K$ in Theorem \ref{theorem7} and the existing bound are replaced by $\mu_B$ and $k\times d$ respectively. Then, the lower bounds of $||\mathbf{x}_{0\backslash\mathbf{S}}||_{2}$ as a function of the number of iteration $l$ are presented in Fig. \ref{lowerboundsx}. It shows that the lower bounds in Theorem \ref{theorem7} and Theorem \ref{theorem8} are smaller than the existing bound for an arbitrary iteration number, which indicates that the conditions required by OLS-type algorithms for reliable reconstruction are more relaxed than that of OMP. Moreover, it implies that BOLS performs better than OLS and MOLS.

\subsection{Empirical simulation tests for the OLS-type algorithms}
This subsection presents empirical tests on the performance of OLS-type algorithms. For each problem setting, we perform $1,000$ independent tests for each algorithm to calculate the frequency of exact recovery. We adopt a Gaussian random matrix $\mathbf{D}\in \mathcal{R}^{M\times N}$ $(M=128$ and $N=256)$ as our measurement matrix with each entry being independently and identically distributed as $\mathcal{N}(0,\frac{1}{M})$.
For each value of $k$, we generate a block $k$-sparse signal, whose support is selected randomly and the nonzero elements independently follow standard Gaussian distribution, i.e., $\mathcal{N}(0,1)$. The recovery is successful if the reconstructed vector is within a certain small Euclidean distance, which is set as $10^{-6}$, of the original vector.

\begin{figure}[htbp]
\centering
\subfigure[]{
\begin{minipage}[t]{0.5\linewidth}
\centering
\includegraphics[width=1.8in]{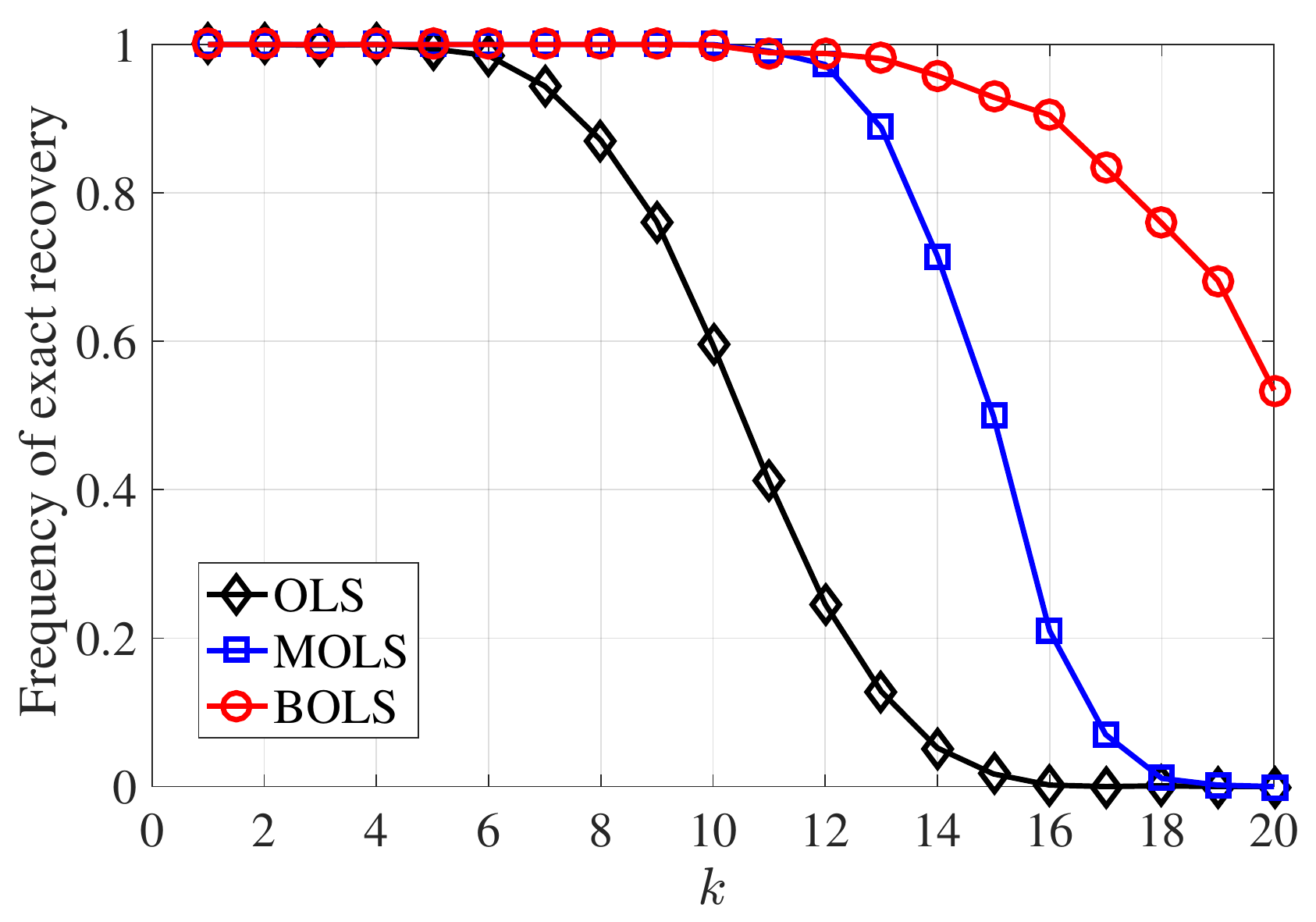}
\end{minipage}%
}%
\subfigure[]{
\begin{minipage}[t]{0.5\linewidth}
\centering
\includegraphics[width=1.8in]{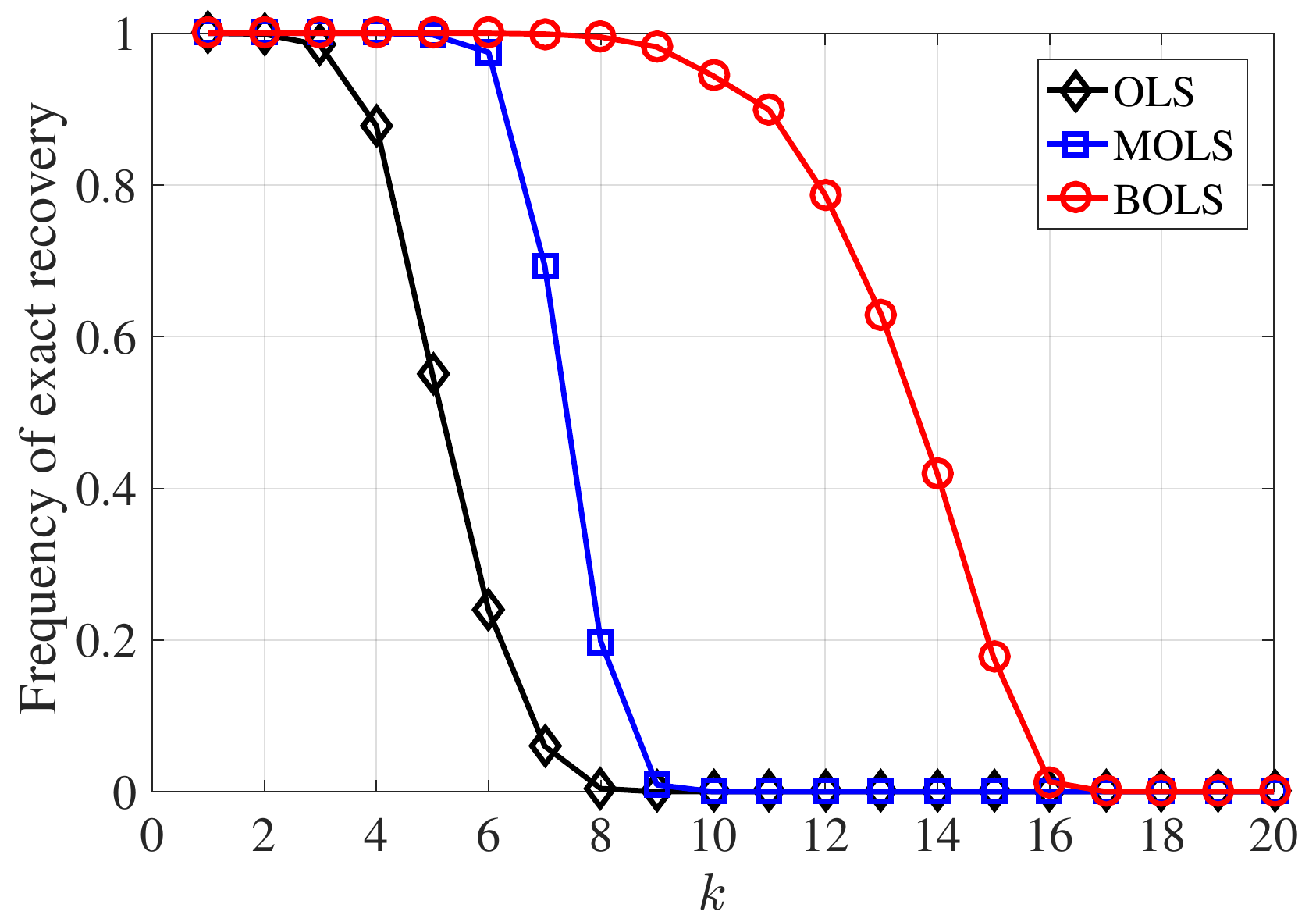}
\end{minipage}%
}%
\centering
\caption{Performance of OLS, MOLS and BOLS with $M=128$, $N=256$, (a) $d=4$; (b) $d=8$.}\label{kdbounds}
\end{figure}

As shown in Fig. \ref{kdbounds}, BOLS outperforms other greedy algorithms, which means that the utilization of the block property exhibits competitive reconstruction performance. When the block length becomes larger, the performance of BOLS is also improved. In Table \ref{table1theoryupper}, we calculate the upper bounds of the reconstructible sparsity levels in Remark~\ref{rmk5} and Remark~\ref{rmk8} for OLS-type algorithms. The coherence parameters are set to be their lower bounds, i.e., $\mu=\frac{1}{\sqrt{M}}$ and $\mu_B=\frac{1}{\sqrt{dM}}$ as given in Section ${\rm \uppercase\expandafter{\romannumeral3}}$-A \cite{1}. It can be seen that the bound $K<\frac{1}{\mu}+1$ is $K<12.3$ in the simulation, which is more relaxed than the sufficient conditions for OLS and MOLS, i.e., $K<8.2$. This means that the Neumann series is convergent in the proof procedures. Evidently, the theoretical thresholds are more pessimistic than the empirical results. However, this negativity is moderated compared with the results in \cite{1}. The theoretical thresholds indicate that the frequencies of exact recovery are close to $100\%$ for OLS and MOLS with $K=8$, for BOLS $(d=4)$ with $kd=18$ and for BOLS $(d=8)$ with $kd=21$. The empirical results show that the frequencies of exact recovery close to $100\%$ for OLS, MOLS, BOLS $(d=4)$ and BOLS $(d=8)$ up to $K=12$, $K=32$, $kd=40$ and $kd=48$ respectively.

\begin{table}[!t]
\renewcommand{\arraystretch}{1.5}
\caption{Upper bounds of reconstructible sparsity levels with the settings in Fig. \ref{kdbounds}}
\label{table1theoryupper}
\centering
\begin{tabular}{|c|c|c|c|c|}
\hline
   & Theoretical condition & Simulation result\\
\hline
OLS        & $K<8.2\Rightarrow K\leq8$ & $K\leq12$ \\
\hline
MOLS    & $K<8.2\Rightarrow K\leq8$ & $K\leq32$\\

\hline
BOLS $(d=4)$      & $kd<18.1\Rightarrow kd\leq18$ & $kd\leq40$ \\
\hline
BOLS $(d=8)$     & $kd<21.2\Rightarrow kd\leq21$ & $kd\leq48$ \\
\hline
$K<1/\mu+1$    & $K<12.3$ & $-$  \\
\hline
\end{tabular}
\end{table}
\section{Conclusion}
In this paper, we study the OLS-type algorithms in reconstructing sparse signals in both noiseless and noisy cases.
We exploit MIP and asymptotic analysis to analyze exact recovery conditions of OLS and MOLS for general sparse signals. Analytical results for BOLS are derived by exploiting the notions of block-coherence and sub-coherence. The various notions of coherence, describing the similarity among matrix columns, are simpler and more intuitive than RIP.
Our derived performance guarantees reveal that the OLS-type algorithms perform reliable recovery if the sparsity level is lower than the MIP-based conditions in noiseless case, or if the power of nonzero coefficients is larger than the conditions related to both MIP and noise variance in noisy case. Due to the sharpness of the mapping normalization factor bound, the developed theoretical results improve the existing ones.


%

\appendices

\section{Proof of Theorem \ref{theorem1}}
\label{proofoftheorem1}
\begin{IEEEproof}
We first use mathematical induction to prove that Theorem \ref{theorem1} is suitable for OLS. Assume that OLS has selected $l$ atoms from $\mathbf{Q}_0$ after $l$ steps $(l\in\{1,\cdots,K-1\})$. Then, we derive a condition to guarantee that the next selected entry is also correct.

Since the expression $||(\mathbf{D}\mathbf{R})^T\mathbf{r}^{l}||_{\infty}$ gives the largest magnitude attained among the inner products, where $\mathbf{R}=f(\mathbf{D})$. In consequence, to see whether the next chosen index corresponds to an atom in $\mathbf{Q}_{0\backslash\mathbf{S}}$, we need to examine whether the following quotient holds:
\begin{equation}\label{olsjudge}
Z(\mathbf{r}^{l},\mathbf{S}^{l})=\frac{||(\overline{\mathbf{Q}}_0\overline{\mathbf{R}}_0)^T\mathbf{r}^{l}||_{\infty}}{||(\mathbf{Q}_{0\backslash\mathbf{S}}\mathbf{R}_{0\backslash\mathbf{S}})^T\mathbf{r}^{l}||_{\infty}}<1.
\end{equation}
Following Tropp's \cite{2} and Soussen's \cite{14} analysis, we know that $\mathbf{r}^{l}\in {\rm span}(\mathbf{Q}_{0\backslash\mathbf{S}})={\rm span}(\mathbf{Q}_{0\backslash\mathbf{S}}\mathbf{R}_{0\backslash\mathbf{S}})$. Therefore,
\begin{equation}\label{DELOLS}
((\mathbf{Q}_{0\backslash\mathbf{S}}\mathbf{R}_{0\backslash\mathbf{S}})^\dag)^T(\mathbf{Q}_{0\backslash\mathbf{S}}\mathbf{R}_{0\backslash\mathbf{S}})^T\mathbf{r}^{l}=\mathbf{r}^{l}.
\end{equation}
Substituting (\ref{DELOLS}) into (\ref{olsjudge}) yields
\begin{align}
&Z(\mathbf{r}^{l},\mathbf{S}^{l})\nonumber\\
=&\frac{||(\overline{\mathbf{Q}}_0\overline{\mathbf{R}}_0)^T((\mathbf{Q}_{0\backslash\mathbf{S}}\mathbf{R}_{0\backslash\mathbf{S}})^\dag)^T(\mathbf{Q}_{0\backslash\mathbf{S}}\mathbf{R}_{0\backslash\mathbf{S}})^T\mathbf{r}^{l}||_{\infty}}{||(\mathbf{Q}_{0\backslash\mathbf{S}}\mathbf{R}_{0\backslash\mathbf{S}})^T\mathbf{r}^{l}||_{\infty}}\nonumber\\
\leq&||(\overline{\mathbf{Q}}_0\overline{\mathbf{R}}_0)^T((\mathbf{Q}_{0\backslash\mathbf{S}}\mathbf{R}_{0\backslash\mathbf{S}})^\dag)^T||_{\infty,\infty}\nonumber\\
=&||(\mathbf{Q}_{0\backslash\mathbf{S}}\mathbf{R}_{0\backslash\mathbf{S}})^\dag(\overline{\mathbf{Q}}_0\overline{\mathbf{R}}_0)||_{1,1}.\label{zlast}
\end{align}

Then, if the following condition holds, i.e.,
\begin{equation}\label{the11111}
Z(\mathbf{r}^{l},\mathbf{S}^{l})\leq||(\mathbf{Q}_{0\backslash\mathbf{S}}\mathbf{R}_{0\backslash\mathbf{S}})^\dag(\overline{\mathbf{Q}}_0\overline{\mathbf{R}}_0)||_{1,1}<1,
\end{equation}
OLS selects a correct atom from $\mathbf{Q}_{0\backslash\mathbf{S}}$.

Next, we prove that the theorem is also valid for MOLS. Omit some similar proof to that of OLS. For MOLS's selecting at least one correct index \cite{10}, it needs to examine whether the following quotient holds:
\begin{equation}\label{MOLSexamine}
Z_M(\mathbf{r}^{l},\mathbf{S}^{l})=\frac{||(\overline{\mathbf{Q}}_0\overline{\mathbf{R}}_0)^T\mathbf{r}^{l}||_{L,1}}{||(\mathbf{Q}_{0\backslash\mathbf{S}}\mathbf{R}_{0\backslash\mathbf{S}})^T\mathbf{r}^{l}||_{\infty}}<L,
\end{equation}
where $||\cdot||_{L,1}$ calculates the absolute sum of $L$ largest elements of its objective. Then, we have
\begin{align}
&Z_M(\mathbf{r}^{l},\mathbf{S}^{l})\nonumber\\
\leq&\frac{L\times||(\overline{\mathbf{Q}}_0\overline{\mathbf{R}}_0)^T\mathbf{r}^{l}||_{\infty}}{||(\mathbf{Q}_{0\backslash\mathbf{S}}\mathbf{R}_{0\backslash\mathbf{S}})^T\mathbf{r}^{l}||_{\infty}}\nonumber\\
=&L\times\frac{||(\overline{\mathbf{Q}}_0\overline{\mathbf{R}}_0)^T((\mathbf{Q}_{0\backslash\mathbf{S}}\mathbf{R}_{0\backslash\mathbf{S}})^\dag)^T(\mathbf{Q}_{0\backslash\mathbf{S}}\mathbf{R}_{0\backslash\mathbf{S}})^T\mathbf{r}^{l}||_{\infty}}{||(\mathbf{Q}_{0\backslash\mathbf{S}}\mathbf{R}_{0\backslash\mathbf{S}})^T\mathbf{r}^{l}||_{\infty}}\nonumber\\
\leq& L\times||(\mathbf{Q}_{0\backslash\mathbf{S}}\mathbf{R}_{0\backslash\mathbf{S}})^\dag(\overline{\mathbf{Q}}_0\overline{\mathbf{R}}_0)||_{1,1}<L,\label{mols2inequality}
\end{align}
and thus the theorem is concluded.
\end{IEEEproof}

\section{Proof of Lemma \ref{lemma3}}
\label{proofoflemma3}
\begin{IEEEproof}
Note that for $i\in\{1,2,\cdots,N\}\backslash \mathbf{S}^l$, $\mathbf{D}_i=\mathbf{P}_{\mathbf{S}^{l}}\mathbf{D}_i+\mathbf{P}_{\mathbf{S}^{l}}^\bot\mathbf{D}_i$. Then,  $||\mathbf{D}_i||_2^2=||\mathbf{P}_{\mathbf{S}^{l}}\mathbf{D}_i||_2^2+||\mathbf{P}_{\mathbf{S}^{l}}^\bot\mathbf{D}_i||_2^2=1$ for normalized matrix $\mathbf{D}$. By rewriting $\mathbf{P}_{\mathbf{S}^{l}}$ and using submultiplicativity of matrix norm, we have
\begin{equation}\label{equalols}
\begin{aligned}
||\mathbf{P}_{\mathbf{S}^{l}}\mathbf{D}_i||_2&=||\mathbf{D}_{\mathbf{S}^{l}}(\mathbf{D}_
{\mathbf{S}^{l}}\mathbf{D}_{\mathbf{S}^{l}})^{-1}\mathbf{D}_{\mathbf{S}^{l}}^T\mathbf{D}_i||_2\\
&\leq\rho(\mathbf{D}_{\mathbf{S}^{l}})\rho((\mathbf{D}_{\mathbf{S}^{l}}^T\mathbf{D}_{\mathbf{S}^{l}})^{-1})||\mathbf{D}_{\mathbf{S}^{l}}^T\mathbf{D}_i||_2.
\end{aligned}
\end{equation}

By exploiting Lemma 2 in \cite{85},
\begin{equation}\label{rounormols}
\rho(\mathbf{D}_{\mathbf{S}^{l}})=
\sqrt{\lambda_{\max}(\mathbf{D}_{\mathbf{S}^{l}}^T\mathbf{D}_{\mathbf{S}^{l}})}
\leq\sqrt{1+(K-1)\mu}.
\end{equation}

Next, we derive the upper bound of the second term in the last inequality of (\ref{equalols}) and we express $\mathbf{D}_{\mathbf{S}^{l}}^T\mathbf{D}_{\mathbf{S}^{l}}$ as $\mathbf{D}_{\mathbf{S}^{l}}^T\mathbf{D}_{\mathbf{S}^{l}}=\mathbf{I}+\mathbf{A}$, where $\mathbf{A}_{ii}=0$ for all $i$. Then, by applying Ger${\rm \check{s}}$gorin's disc theorem, we obtain
\begin{equation}\label{rouA}
\rho(\mathbf{A})\leq(l-1)\mu\leq(K-1)\mu<1.
\end{equation}
Using Neumann series expansion (Corollary 5.6.16 \cite{3}), we have $(\mathbf{I}+\mathbf{A})^{-1}=\sum_{k=0}^{\infty}(-\mathbf{A})^k$. Therefore,
\begin{equation}\label{midols}
\begin{aligned}
\rho((\mathbf{D}_{\mathbf{S}^{l}}^T\mathbf{D}_{\mathbf{S}_{l}})^{-1})=&
\rho\Big(\sum_{k=0}^\infty(-\mathbf{A})^k\Big)
\leq\sum^\infty_{k=0}\big(\rho(\mathbf{A})^k\big)\\
=&\frac{1}{1-\rho(\mathbf{A})}
\leq\frac{1}{1-(K-1)\mu}.
\end{aligned}
\end{equation}

Moreover, the upper bound of $||\mathbf{D}_{\mathbf{S}^{l}}^T\mathbf{D}_i||_2$ is calculated by
\begin{equation}\label{thirdols}
||\mathbf{D}_{\mathbf{S}^{l}}^T\mathbf{D}_i||_2\leq\sqrt{l\mu^2}\leq\sqrt{K\mu^2}.
\end{equation}

Combining (\ref{rounormols}), (\ref{midols}) and (\ref{thirdols}) yields
\begin{equation}\label{upper}
\begin{aligned}
||\mathbf{P}_{\mathbf{S}^{l}}\mathbf{D}_i||_2&\leq\frac{\sqrt{1+(K-1)\mu}\sqrt{K\mu^2}}{1-(K-1)\mu}.
\end{aligned}
\end{equation}

Finally, we obtain
\begin{equation}\label{lastinols}
\begin{aligned}
1&\overset{(a)}{\geq}||\mathbf{P}_{\mathbf{S}^{l}}^\bot \mathbf{D}_i||_2=\sqrt{1-||\mathbf{P}_{\mathbf{S}^{l}}\mathbf{D}_i||_2^2}\\
&\geq\sqrt{1-\bigg(\frac{\sqrt{1+(K-1)\mu}\sqrt{K\mu^2}}{1-(K-1)\mu}\bigg)^2}.
\end{aligned}
\end{equation}
The reason why $(a)$ in (\ref{lastinols}) holds is that in the first iteration, $\mathbf{S}^0=\mathbf{\emptyset}$, and thus $||\mathbf{P}^\bot_{\mathbf{S}^{0}}\mathbf{D}_i||_2=||\mathbf{D}_i||_2=1$. For $l>1$, $||\mathbf{P}_{\mathbf{S}^{l}}^\bot \mathbf{D}_i||_2<||\mathbf{D}_i||_2=1$.
\end{IEEEproof}

\section{Proof of Lemma \ref{lemma4}}
\label{proofoflemma4}
\begin{IEEEproof}
Define $\mathbf{\mathbf{a}}^j=\mathbf{B}^T_{\bullet\backslash j}\mathbf{B}_j\in \mathcal{R}^{K-1}$ $(j\in\{1,2,\cdots,K\})$, where $\mathbf{B}_{\bullet\backslash j}=\mathbf{B}\backslash \mathbf{B}_j$. Then, according to the central limit theorem, when $M$ is sufficiently large, the entries of $\mathbf{a}^j$ are independently and identically distributed as $\mathcal{N}(0,\frac{1}{M})$. Meanwhile, the random variable $\mathcal{A}^j = M||\mathbf{a}^j||_2^2\sim\chi^2_{K-1}$. Therefore,
\begin{equation}\label{k1half-normaldis}
\begin{aligned}
&{\rm P}\Big\{||\mathbf{a}^j||_1\leq\frac{(K-\mathcal{T})\tau}{2}\Big\}\\
\geq&{\rm P}\Big\{\sqrt{K-1}||\mathbf{a}^j||_2\leq\frac{(K-\mathcal{T})\tau}{2}\Big\}\\
=&{\rm P}\Big\{M||\mathbf{a}^j||^2_2\leq\frac{M(K-\mathcal{T})^2\tau^2}{4(K-1)}\Big\}.
\end{aligned}
\end{equation}

Moreover, the Lemma 4 in \cite{87} indicates that for any $\mathcal{C}>0$,
\begin{equation}\label{lemma4incite}
{\rm P}\{\mathcal{A}^j>(1+\mathcal{C})(K-1)\}\leq\frac{e^{-\frac{K-1}{2}(\mathcal{C}-\log(1+\mathcal{C}))}}{\mathcal{C}\sqrt{\pi(K-1)}}.
\end{equation}

Hence,
\begin{equation}\label{thelastprob}
\begin{aligned}
&{\rm P}\Big\{||\mathbf{B}^T\mathbf{B}-\mathbf{I}||_{1,1}\leq \frac{(K-\mathcal{T})\tau}{2}\Big\}\\
=&1-{\rm P}\Big\{\bigcup_{j=1}^K\Big(||\mathbf{a}^j||_1>\frac{(K-\mathcal{T})\tau}{2}\Big)\Big\}\\
\geq&1-\sum_{j=1}^{K}{\rm P}\Big\{||\mathbf{a}^j||_1>\frac{(K-\mathcal{T})\tau}{2}\Big\}\\
\geq&1-\frac{Ke^{-\frac{K-1}{2}(\mathcal{C}-\log(1+\mathcal{C}))}}{\mathcal{C}\sqrt{\pi(K-1)}},
\end{aligned}
\end{equation}
where $\mathcal{C}=\frac{M(K-\mathcal{T})^2\tau^2}{4(K-1)^2}-1$.
\end{IEEEproof}

\section{Proof of Theorem \ref{theorem2}}
\label{proofoftheorem2}
\begin{IEEEproof}
By exploiting submultiplicativity,
\begin{align}
&||(\mathbf{Q}_{0\backslash\mathbf{S}}\mathbf{R}_{0\backslash\mathbf{S}})^\dag(\overline{\mathbf{Q}}_0\overline{\mathbf{R}}_0)||_{1,1}  \nonumber\\
=&\max_{\overline{\mathbf{G}}_i}{||(\mathbf{Q}_{0\backslash\mathbf{S}}\mathbf{R}_{0\backslash\mathbf{S}})^\dag\overline{\mathbf{G}}_i||_1}\nonumber\\
=&\max_{\overline{\mathbf{G}}_i}{||((\mathbf{Q}_{0\backslash\mathbf{S}}\mathbf{R}_{0\backslash\mathbf{S}})^T\mathbf{Q}_{0\backslash\mathbf{S}}\mathbf{R}_{0\backslash\mathbf{S}})^{-1}(\mathbf{Q}_{0\backslash\mathbf{S}}\mathbf{R}_{0\backslash\mathbf{S}})^T\overline{\mathbf{G}}_i||_1}\nonumber\\
\leq&||((\mathbf{Q}_{0\backslash\mathbf{S}}\mathbf{R}_{0\backslash\mathbf{S}})^T\mathbf{Q}_{0\backslash\mathbf{S}}\mathbf{R}_{0\backslash\mathbf{S}})^{-1}||_{1,1}\nonumber\\
&\times\max_{\overline{\mathbf{G}}_i}||(\mathbf{Q}_{0\backslash\mathbf{S}}\mathbf{R}_{0\backslash\mathbf{S}})^T\overline{\mathbf{G}}_i||_1, \label{writeout}
\end{align}
where $\overline{\mathbf{G}}_i$ represents the $i$-th column of $\overline{\mathbf{Q}}_0\overline{\mathbf{R}}_0$.
Now let us bound the term  $||((\mathbf{Q}_{0\backslash\mathbf{S}}\mathbf{R}_{0\backslash\mathbf{S}})^T\mathbf{Q}_{0\backslash\mathbf{S}}\mathbf{R}_{0\backslash\mathbf{S}})^{-1}||_{1,1}$ in the last inequality of (\ref{writeout}).
Let the off-diagonal part of $\mathbf{Q}_{0\backslash\mathbf{S}}^T\mathbf{Q}_{0\backslash\mathbf{S}}$ be $\mathbf{A}$, i.e., $\mathbf{Q}_{0\backslash\mathbf{S}}^T\mathbf{Q}_{0\backslash\mathbf{S}}=\mathbf{I}+\mathbf{A}$.
Then, according to Lemma~\ref{lemma4}, we have
\begin{equation}\label{accordingtole4}
\begin{aligned}
||\mathbf{A}||_{1,1}\leq\frac{(K-l-\mathcal{T})\mu}{2}\leq\frac{(K-\mathcal{T})\mu}{2}.
\end{aligned}
\end{equation}
Now the assumption (\ref{theorem2mian2}) implies $\frac{(K-\mathcal{T})\mu}{2}<1$. Then,
by exploiting Neumann series expansion,
\begin{equation}\label{neumannols}
\begin{aligned}
||(\mathbf{Q}_{0\backslash\mathbf{S}}^T\mathbf{Q}_{0\backslash\mathbf{S}})^{-1}||_{1,1}=&\Big|\Big|\sum_{k=0}^\infty(-\mathbf{A})^k\Big|\Big|_{1,1}\leq\frac{1}{1-\frac{K\mu}{2}+\frac{\mathcal{T}\mu}{2}}.
\end{aligned}
\end{equation}

Observe that
\begin{align}
&||((\mathbf{Q}_{0\backslash\mathbf{S}}\mathbf{R}_{0\backslash\mathbf{S}})^T\mathbf{Q}_{0\backslash\mathbf{S}}\mathbf{R}_{0\backslash\mathbf{S}})^{-1}||_{1,1}\nonumber\\
\leq&||(\mathbf{R}_{0\backslash\mathbf{S}})^{-1}||_{1,1}||(\mathbf{Q}_{0\backslash\mathbf{S}}^T\mathbf{Q}_{0\backslash\mathbf{S}})^{-1}||_{1,1}||(\mathbf{R}_{0\backslash\mathbf{S}})^{-1}||_{1,1}\nonumber\\
\leq&||(\mathbf{Q}_{0\backslash\mathbf{S}}^T\mathbf{Q}_{0\backslash\mathbf{S}})^{-1}||_{1,1}\leq\frac{1}{1-\frac{K\mu}{2}+\frac{\mathcal{T}\mu}{2}}.\label{meanwhile}
\end{align}

On the other hand, by the definition of matrix coherence, it is straightforward to obtain
\begin{equation}\label{defcoherence}
\max_{\overline{\mathbf{G}}_i}||(\mathbf{Q}_{0\backslash\mathbf{S}}\mathbf{R}_{0\backslash\mathbf{S}})^T\overline{\mathbf{G}}_i||_1\leq K\mathcal{T}\mu.
\end{equation}
Combining (\ref{meanwhile}) and (\ref{defcoherence}), we get
\begin{equation}\label{olsfinally}
||(\mathbf{Q}_{0\backslash\mathbf{S}}\mathbf{R}_{0\backslash\mathbf{S}})^\dag(\overline{\mathbf{Q}}_0\overline{\mathbf{R}}_0)||_{1,1}\leq\frac{2K\mathcal{T}\mu}{2-(K-\mathcal{T})\mu}<1.
\end{equation}

By writing out $\mathcal{T}$, we obtain the following cubic inequality,
\begin{equation}\label{ols cubic inequality}
\alpha K^3+\beta K^2+\gamma K+\delta<0,
\end{equation}
where $\alpha$, $\beta$, $\gamma$ and $\delta$ are given in the description of Theorem~\ref{theorem2}.
Finally, by exploiting Cardano formula to solve the valid solution of the inequality (\ref{ols cubic inequality}) in real domain, the proof is concluded.
\end{IEEEproof}

\section{Proof of Lemma \ref{lemma2}}
\label{proofoflemma2}
\begin{IEEEproof}
According to submultiplicativity of matrix norm,
\begin{equation}\label{equal}
\begin{aligned}
||\mathbf{P}_{\mathbf{S}^{l}}\mathbf{D}_i||_2&=||\mathbf{D}_{\mathbf{S}^{l}}(\mathbf{D}_
{\mathbf{S}^{l}}\mathbf{D}_{\mathbf{S}^{l}})^{-1}\mathbf{D}_{\mathbf{S}^{l}}^T\mathbf{D}_i||_2\\
&\leq\rho(\mathbf{D}_{\mathbf{S}^{l}})\rho((\mathbf{D}_{\mathbf{S}^{l}}^T\mathbf{D}_{\mathbf{S}^{l}})^{-1})||\mathbf{D}_{\mathbf{S}^{l}}^T\mathbf{D}_i||_2.
\end{aligned}
\end{equation}

Similar to that in Appendix \ref{proofoflemma3}, we have
\begin{equation}\label{simialrtoappen}
\rho(\mathbf{D}_{\mathbf{S}^{l}})
\leq\sqrt{1+(kd-1)\mu}
\end{equation}
and
\begin{equation}\label{simialrtoappen2}
||\mathbf{D}_{\mathbf{S}^{l}}^T\mathbf{D}_i||_2\leq\sqrt{l\mu^2}\leq\sqrt{kd\mu^2}.
\end{equation}

Then, it remains to derive the upper bound of the second term in the last inequality of (\ref{equal}) and we express $\mathbf{D}_{\mathbf{S}^{l}}^T\mathbf{D}_{\mathbf{S}^{l}}$ as $\mathbf{D}_{\mathbf{S}^{l}}^T\mathbf{D}_{\mathbf{S}^{l}}=\mathbf{I}+\mathbf{A}$, where $\mathbf{A}$ is an $ld\times ld$ matrix with blocks $\mathbf{A}[l,r]$ of size $d\times d$ such that $\mathbf{A}_{ii}=0$ for all $i$ because of the normalized matrix $\mathbf{D}$. Since $\mathbf{A}[l,r]=\mathbf{D}^T_{\mathbf{S}^l}[l]\mathbf{D}_{\mathbf{S}^l}[r]$ for all $l\neq r$, and $\mathbf{A}[r,r]=\mathbf{D}^T_{\mathbf{S}^l}[r]\mathbf{D}_{\mathbf{S}^l}[r]-\mathbf{I}$, we have
\begin{align}
\rho(\mathbf{A})&\leq(d-1)\nu+(l-2)d\mu\leq(d-1)\nu+(k-1)d\mu\nonumber\\
&\leq(kd-1)\mu<1.\label{rouA}
\end{align}

Therefore, $(\mathbf{I}+\mathbf{A})^{-1}=\sum_{k=0}^{\infty}(-\mathbf{A})^k$ and we obtain
\begin{equation}\label{mid}
\begin{aligned}
\rho((\mathbf{D}_{\mathbf{S}^{l}}^T\mathbf{D}_{\mathbf{S}^{l}})^{-1})=&
\rho(\sum_{k=0}^\infty(-\mathbf{A})^k)\leq\sum^\infty_{k=0}(\rho(\mathbf{A})^k)\\
=&\frac{1}{1-\rho(\mathbf{A})}\\
\leq&\frac{1}{1-(d-1)\nu-(k-1)d\mu}.
\end{aligned}
\end{equation}

Hence,
\begin{equation}\label{upper}
\begin{aligned}
||\mathbf{P}_{\mathbf{S}^{l}}\mathbf{D}_i||^2_2&\leq\frac{kd\mu^2+kd(kd-1)\mu^3}{(1-(d-1)\nu-(k-1)d\mu)^2}.
\end{aligned}
\end{equation}

Finally, we have
\begin{equation}\label{lastin}
\begin{aligned}
1&\geq||\mathbf{P}_{\mathbf{S}^{l}}^\bot \mathbf{D}_i||_2=\sqrt{1-||\mathbf{P}_{\mathbf{S}^{l}}\mathbf{D}_i||_2^2}\\
&\geq\sqrt{1-\frac{kd\mu^2+kd(kd-1)\mu^3}{(1-(d-1)\nu-(k-1)d\mu)^2}}.
\end{aligned}
\end{equation}
\end{IEEEproof}

\section{Proof of Proposition \ref{proposition1}}
\label{proofofproposition1}
\begin{IEEEproof}
From the geometric properties of a right triangle, we get
\begin{equation}\label{PLUS}
\begin{aligned}
&||\mathbf{P}^\bot_{\mathbf{S}^{l}\cup \{(j_B-1)d+1,\cdots,j_Bd\}}\mathbf{y}||^2_2\\
+&||\mathbf{P}_{\mathbf{S}^{l}\cup \{(j_B-1)d+1,\cdots,j_Bd\}}\mathbf{y}||^2_2=||\mathbf{y}||_2^2.
\end{aligned}
\end{equation}
Therefore, support selection step in BOLS is altered to
\begin{equation}\label{puls2}
i^{l+1}_B=\mathop{\arg\max}\limits_{j_B\in\{1,..,N_B\}\backslash\mathbf{S}_B^{l}}||\mathbf{P}_{\mathbf{S}^{l}\cup \{(j_B-1)d+1,\cdots,j_Bd\}}\mathbf{y}||_2^2.
\end{equation}
Without loss of generality, we assume the $i_B$-th block contains the indices $\{1,\cdots,d\}$, where $d\geq2$.
Then, thanks to the decomposition characteristics \cite{20} of
$||\mathbf{P}_{\mathbf{S}^{l}\cup \{1,\cdots,d\}}\mathbf{y}||_2^2$,
we have
\begin{equation}\label{decomp}
\begin{aligned}
&||\mathbf{P}_{\mathbf{S}^{l}\cup \{1,\cdots,d\}}\mathbf{y}||_2^2\\
=&||\mathbf{P}_{\mathbf{S}^{l}\cup\{1,\cdots,d-1\}}\mathbf{y}||^2_2
+\Bigg(\frac{|<\mathbf{D}_d,\mathbf{r}^{l}>|}{||\mathbf{P}^\bot_{\mathbf{S}^{l}\cup\{1,\cdots,d-1\}}\mathbf{D}_d||_2}\Bigg)^2\\
\overset{(a)}{=}&||\mathbf{P}_{\mathbf{S}^{l}}\mathbf{y}||^2_2+\Bigg(\frac{|<\mathbf{D}_1,\mathbf{r}^{l}>|}{||\mathbf{P}^\bot_{\mathbf{S}^{l}}\mathbf{D}_1||_2}\Bigg)^2\\
&+\sum^d_{j=2}\Bigg(\frac{|<\mathbf{D}_j,\mathbf{r}^{l}>|}{||\mathbf{P}^\bot_{\mathbf{S}^{l}\cup\{1,\cdots,j-1\}}\mathbf{D}_j||_2}\Bigg)^2,
\end{aligned}
\end{equation}
where equation $(a)$ is derived by writing out $||\mathbf{P}_{\mathbf{S}^{l}\cup\{1,\cdots,d-1\}}\mathbf{y}||^2_2$ and we obtain (\ref{ctri}).
\end{IEEEproof}

\section{Proof of Theorem \ref{theorem5}}
\label{proofoftheorem5}
\begin{IEEEproof}
Based on Proposition~\ref{proposition1}, the atom selection indicator of the BOLS algorithm is given by
\begin{equation}\label{2-wuqiong}
\begin{aligned}
||(\mathbf{D}_{\bullet\backslash \mathbf{S}}\mathbf{R}_{\bullet\backslash \mathbf{S}})^T\mathbf{r}^{l}||_{2,\infty},
\end{aligned}
\end{equation}
where $\mathbf{D}_{\bullet\backslash \mathbf{S}} =\mathbf{D}\backslash \mathbf{D}_{\mathbf{S}^l} \in\mathcal{R}^{M\times (N-ld)}$, and $\mathbf{R}_{\bullet\backslash \mathbf{S}} \in\mathcal{R}^{(N-ld)\times (N-ld)}$ is given as follows:
\begin{equation}\label{r000000}
\mathbf{R}_{\bullet\backslash \mathbf{S}}=\left[
\begin{matrix}

   \mathbf{R}_{0\backslash \mathbf{S}} & \mathbf{0} \\

   \mathbf{0} &\overline{\mathbf{R}}_{0}\\

\end{matrix}
\right].
\end{equation}

Similar to the assumption in Appendix \ref{proofoftheorem1}, to see whether the next chosen index corresponds to an atom in $\mathbf{Q}_{0\backslash\mathbf{S}}$, we need to examine whether the following quotient holds:
\begin{equation}\label{THE1PROOF}
Z(\mathbf{r}^{l},\mathbf{S}^{l})=\frac{||(\overline{\mathbf{Q}}_0\overline{\mathbf{R}}_0)^T\mathbf{r}^{l}||_{2,\infty}}{||(\mathbf{Q}_{0\backslash\mathbf{S}}\mathbf{R}_{0\backslash\mathbf{S}})^T\mathbf{r}^{l}||_{2,\infty}}<1.
\end{equation}

Since $\mathbf{Q}_{0\backslash\mathbf{S}}\mathbf{R}_{0\backslash\mathbf{S}}(\mathbf{Q}_{0\backslash\mathbf{S}}\mathbf{R}_{0\backslash\mathbf{S}})^\dag$ is Hermitian, we have
\begin{equation}\label{DEL}
((\mathbf{Q}_{0\backslash\mathbf{S}}\mathbf{R}_{0\backslash\mathbf{S}})^\dag)^T(\mathbf{Q}_{0\backslash\mathbf{S}}\mathbf{R}_{0\backslash\mathbf{S}})^T\mathbf{r}^{l}=\mathbf{r}^{l}.
\end{equation}
Substituting (\ref{DEL}) into (\ref{THE1PROOF}) yields
\begin{equation}\label{zlast}
\begin{aligned}
Z(\mathbf{r}^{l},\mathbf{S}^{l})&\leq\rho_r((\overline{\mathbf{Q}}_0\overline{\mathbf{R}}_0)^T((\mathbf{Q}_{0\backslash\mathbf{S}}\mathbf{R}_{0\backslash\mathbf{S}})^\dag)^T)\\
&=\rho_c((\mathbf{Q}_{0\backslash\mathbf{S}}\mathbf{R}_{0\backslash\mathbf{S}})^\dag(\overline{\mathbf{Q}}_0\overline{\mathbf{R}}_0)).
\end{aligned}
\end{equation}
The following proof of BOLS in selecting a new block index in each step is similar to that in Section ${\rm\uppercase\expandafter{\romannumeral5}}$-A \cite{1}. Finally, we conclude the proof.
\end{IEEEproof}

\section{Proof of Theorem \ref{theorem6}}
\label{proofoftheorem6}
\begin{IEEEproof}
By using submultiplicativity, we have
\begin{align}
&\rho_c((\mathbf{Q}_{0\backslash\mathbf{S}}\mathbf{R}_{0\backslash\mathbf{S}})^\dag(\overline{\mathbf{Q}}_0\overline{\mathbf{R}}_0))\nonumber\\
=&\rho_c(((\mathbf{Q}_{0\backslash\mathbf{S}}\mathbf{R}_{0\backslash\mathbf{S}})^T\mathbf{Q}_{0\backslash\mathbf{S}}\mathbf{R}_{0\backslash\mathbf{S}})^{-1}(\mathbf{Q}_{0\backslash\mathbf{S}}\mathbf{R}_{0\backslash\mathbf{S}})^T\overline{\mathbf{Q}}_0\overline{\mathbf{R}}_0)\nonumber\\
\leq&\rho_c(((\mathbf{Q}_{0\backslash\mathbf{S}}\mathbf{R}_{0\backslash\mathbf{S}})^T\mathbf{Q}_{0\backslash\mathbf{S}}\mathbf{R}_{0\backslash\mathbf{S}})^{-1})\nonumber\\
&\times\rho_c((\mathbf{Q}_{0\backslash\mathbf{S}}\mathbf{R}_{0\backslash\mathbf{S}})^T\overline{\mathbf{Q}}_0\overline{\mathbf{R}}_0).\label{inverse}
\end{align}

Denote the off-diagonal part of $\mathbf{Q}_{0\backslash\mathbf{S}}^T\mathbf{Q}_{0\backslash\mathbf{S}}$ as $\mathbf{A}$. Then, due to Lemma \ref{lemma5},
\begin{equation}\label{duetolemma5}
\rho_c(\mathbf{A})\leq\frac{(k-l-\mathcal{T}_B)d\mu_B}{2}\leq\frac{(k-\mathcal{T}_B)d\mu_B}{2}.
\end{equation}

The result (\ref{theorem6mian}) now indicates that $\frac{(k-\mathcal{T}_B)d\mu_B}{2}<1$ and thus $\rho_c(\mathbf{A})<1$, which leading to the following Neumann series expansion,
\begin{align}
\rho_c((\mathbf{Q}_{0\backslash\mathbf{S}}^T\mathbf{Q}_{0\backslash\mathbf{S}})^{-1})=&\rho_c\Big(\sum_{k=0}^\infty(-\mathbf{A})^k\Big)\leq\sum_{k=0}^\infty\rho_c\Big((-\mathbf{A})^k\Big)\nonumber\\
\leq&\frac{1}{1-\frac{(k-\mathcal{T}_B)d\mu_B}{2}}.\label{neumann}
\end{align}

Therefore,
\begin{equation}\label{thereforene}
\begin{aligned}
&\rho_c(((\mathbf{Q}_{0\backslash\mathbf{S}}\mathbf{R}_{0\backslash\mathbf{S}})^T\mathbf{Q}_{0\backslash\mathbf{S}}\mathbf{R}_{0\backslash\mathbf{S}})^{-1})\\
\leq&\rho_c(\mathbf{R}^{-1}_{0\backslash\mathbf{S}})\rho_c((\mathbf{Q}^T_{0\backslash\mathbf{S}}\mathbf{Q}_{0\backslash\mathbf{S}})^{-1})\rho_c((\mathbf{R}^T_{0\backslash\mathbf{S}})^{-1})\\
\leq&\rho_c((\mathbf{Q}_{0\backslash\mathbf{S}}^T\mathbf{Q}_{0\backslash\mathbf{S}})^{-1})\\
\leq&\frac{2}{2-(k-\mathcal{T}_B)d\mu_B}.
\end{aligned}
\end{equation}

Note that
\begin{equation}\label{secondlastone}
\begin{aligned}
&\rho_c((\mathbf{Q}_{0\backslash\mathbf{S}}\mathbf{R}_{0\backslash\mathbf{S}})^T(\overline{\mathbf{Q}}_0\overline{\mathbf{R}}_0))\\
=&\rho_c(\mathbf{R}_{0\backslash\mathbf{S}}^T\mathbf{Q}_{0\backslash\mathbf{S}}^T\overline{\mathbf{Q}}_0\overline{\mathbf{R}}_0)\\
\leq&\rho_c(\mathbf{R}_{0\backslash\mathbf{S}}^T)\rho_c(\mathbf{Q}_{0\backslash\mathbf{S}}^T\overline{\mathbf{Q}}_0)\rho_c(\overline{\mathbf{R}}_0)\\
\leq&\mathcal{T}_Bkd\mu_B.
\end{aligned}
\end{equation}
Combining (\ref{thereforene}) and (\ref{secondlastone}), we have
\begin{equation}\label{final}
\begin{aligned}
\rho_c((\mathbf{Q}_{0\backslash\mathbf{S}}\mathbf{R}_{0\backslash\mathbf{S}})^\dag(\overline{\mathbf{Q}}_0\overline{\mathbf{R}}_0))&\leq\frac{2\mathcal{T}_Bkd\mu_B}{2-(k-\mathcal{T}_B)d\mu_B}\\
&<1.
\end{aligned}
\end{equation}

Simplify (\ref{final}) and we get a cubic inequality with respect to $kd$,
\begin{equation}\label{cubic inequality1}
\alpha_B(kd)^3+\beta_B(kd)^2+\gamma_Bkd+\delta_B<0,
\end{equation}
where $\alpha_B$, $\beta_B$, $\gamma_B$ and $\delta_B$ are given in the description of the theorem.
Finally, the proof is concluded by exploiting Cardano formula to calculate the solution of (\ref{cubic inequality1}) in real domain.
\end{IEEEproof}

\section{Proof of Theorem \ref{theorem7}}
\label{proofoftheorem7}
\begin{IEEEproof}
We first prove the theorem for OLS.
It is clear that a sufficient condition for OLS to select a correct atom in the $(l+1)$-th iteration is
\begin{equation}\label{asufficientcondition}
||(\mathbf{Q}_{0\backslash\mathbf{S}}\mathbf{R}_{0\backslash\mathbf{S}})^T\mathbf{r}^{l}||_{\infty}>||(\overline{\mathbf{Q}}_0\overline{\mathbf{R}}_0)^T\mathbf{r}^{l}||_{\infty}.
\end{equation}

By using the residual definition (\ref{residual}), a sufficient condition for  (\ref{asufficientcondition}) is
\begin{equation}\label{becomses}
\begin{aligned}
&||(\mathbf{Q}_{0\backslash\mathbf{S}}\mathbf{R}_{0\backslash\mathbf{S}})^T\mathbf{s}^{l}||_{\infty}-||(\mathbf{Q}\mathbf{R})^T\mathbf{n}^{l}||_{\infty}\\
>&||(\overline{\mathbf{Q}}_0\overline{\mathbf{R}}_0)^T\mathbf{s}^{l}||_{\infty}+||(\mathbf{Q}\mathbf{R})^T\mathbf{n}^{l}||_{\infty},
\end{aligned}
\end{equation}
which implies
\begin{equation}\label{implies}
\begin{aligned}
&||(\mathbf{Q}_{0\backslash\mathbf{S}}\mathbf{R}_{0\backslash\mathbf{S}})^T\mathbf{s}^{l}||_{\infty}-||(\overline{\mathbf{Q}}_0\overline{\mathbf{R}}_0)^T\mathbf{s}^{l}||_{\infty}
>2||(\mathbf{Q}\mathbf{R})^T\mathbf{n}^{l}||_{\infty}.
\end{aligned}
\end{equation}

Following the Lemma 4 in \cite{85} and (\ref{olsksteps}), we have
\begin{equation}\label{lemma485}
\begin{aligned}
&||(\mathbf{Q}_{0\backslash\mathbf{S}}\mathbf{R}_{0\backslash\mathbf{S}})^T\mathbf{s}^{l}||_{\infty}-||(\overline{\mathbf{Q}}_0\overline{\mathbf{R}}_0)^T\mathbf{s}^{l}||_{\infty}\\
\geq&(1-||(\mathbf{Q}_{0\backslash\mathbf{S}}\mathbf{R}_{0\backslash\mathbf{S}})^\dag(\overline{\mathbf{Q}}_0\overline{\mathbf{R}}_0)||_{1,1})||(\mathbf{Q}_{0\backslash\mathbf{S}}\mathbf{R}_{0\backslash\mathbf{S}})^T\mathbf{s}^{l}||_{\infty}.
\end{aligned}
\end{equation}

Combining (\ref{implies}), (\ref{lemma485}) and Lemma \ref{lemma6}, the condition
\begin{equation}\label{combineyields}
||(\mathbf{Q}_{0\backslash\mathbf{S}}\mathbf{R}_{0\backslash\mathbf{S}})^T\mathbf{s}^{l}||_{\infty}>\frac{2(2-(K-\mathcal{T})\mu)||(\mathbf{Q}\mathbf{R})^T\mathbf{n}^{l}||_{\infty}}{2-(K-\mathcal{T})\mu-2K\mathcal{T}\mu}
\end{equation}
ensures that OLS selects a correct atom in the $(l+1)$-th iteration.

Note that
\begin{equation}\label{notethat}
\begin{aligned}
&||(\mathbf{Q}_{0\backslash\mathbf{S}}\mathbf{R}_{0\backslash\mathbf{S}})^T\mathbf{s}^{l}||_{\infty}\\
=&||(\mathbf{Q}_{0\backslash\mathbf{S}}\mathbf{R}_{0\backslash\mathbf{S}})^T(\mathbf{I}-\mathbf{P}_{\mathbf{S}^l})\mathbf{D}_{0\backslash\mathbf{S}}\mathbf{x}_{0\backslash\mathbf{S}}||_{\infty}\\
\geq&\frac{||(\mathbf{Q}_{0\backslash\mathbf{S}}\mathbf{R}_{0\backslash\mathbf{S}})^T(\mathbf{I}-\mathbf{P}_{\mathbf{S}^l})\mathbf{D}_{0\backslash\mathbf{S}}\mathbf{x}_{0\backslash\mathbf{S}}||_{2}}{\sqrt{K-l}}\\
\geq&\frac{(1-(K-1)\mu)||\mathbf{x}_{0\backslash\mathbf{S}}||_{2}}{\sqrt{K-l}},
\end{aligned}
\end{equation}
where the last inequality follows from Lemmas 2 and 5 in \cite{85}. Combining (\ref{combineyields}) and (\ref{notethat}) yields
\begin{equation}\label{thelastth}
||\mathbf{x}_{0\backslash\mathbf{S}}||_{2}>\frac{2\sqrt{K-l}(2-(K-\mathcal{T})\mu)||(\mathbf{Q}\mathbf{R})^T\mathbf{n}^{l}||_{\infty}}{(2-(K-\mathcal{T})\mu-2K\mathcal{T}\mu)(1-(K-1)\mu)}.
\end{equation}

Then, using Lemma 5.1 in \cite{86}, we obtain that (\ref{remainingentries})
guarantees OLS for selecting a correct atom in the $(l+1)$-th iteration with the probability at least $1-\frac{1}{M}$.

For MOLS algorithm, the sufficient condition for selecting at least one correct atom is
\begin{equation}\label{MOLSselection}
\begin{aligned}
L||(\mathbf{Q}_{0\backslash\mathbf{S}}\mathbf{R}_{0\backslash\mathbf{S}})^T\mathbf{r}^{l}||_{\infty}>&||(\overline{\mathbf{Q}}_0\overline{\mathbf{R}}_0)^T\mathbf{r}^{l}||_{L,1}\\
\geq& L||(\overline{\mathbf{Q}}_0\overline{\mathbf{R}}_0)^T\mathbf{r}^{l}||_{\infty},
\end{aligned}
\end{equation}
where $||\cdot||_{L,1}$ is defined in Appendix \ref{proofoftheorem1}. Then, by eliminating $L$ on both sides of the inequality (\ref{MOLSselection}), the proof is concluded for MOLS by using the similar procedures as described above.
\end{IEEEproof}

\section{Proof of Theorem \ref{theorem8}}
\label{proofoftheorem8}
\begin{IEEEproof}
A sufficient condition for BOLS to select a correct block in the $(l+1)$-th iteration is
\begin{equation}\label{asufficientconditionbols}
||(\mathbf{Q}_{0\backslash\mathbf{S}}\mathbf{R}_{0\backslash\mathbf{S}})^T\mathbf{r}^{l}||_{2,\infty}>||(\overline{\mathbf{Q}}_0\overline{\mathbf{R}}_0)^T\mathbf{r}^{l}||_{2,\infty}.
\end{equation}

Similar to the proof in Appendix I,
\begin{equation}\label{impliesbols}
\begin{aligned}
&||(\mathbf{Q}_{0\backslash\mathbf{S}}\mathbf{R}_{0\backslash\mathbf{S}})^T\mathbf{s}^{l}||_{2,\infty}-||(\overline{\mathbf{Q}}_0\overline{\mathbf{R}}_0)^T\mathbf{s}^{l}||_{2,\infty}\\
>&2||(\mathbf{Q}\mathbf{R})^T\mathbf{n}^{l}||_{2,\infty}
\end{aligned}
\end{equation}
is the sufficient condition for (\ref{asufficientconditionbols}).
By using Lemma 4 in \cite{85} and (\ref{sufficient condition}), we have
\begin{equation}\label{lemma485bols}
\begin{aligned}
&||(\mathbf{Q}_{0\backslash\mathbf{S}}\mathbf{R}_{0\backslash\mathbf{S}})^T\mathbf{s}^{l}||_{2,\infty}-||(\overline{\mathbf{Q}}_0\overline{\mathbf{R}}_0)^T\mathbf{s}^{l}||_{2,\infty}\\
\geq&(1-\rho_c((\mathbf{Q}_{0\backslash \mathbf{S}}\mathbf{R}_{0\backslash \mathbf{S}})^\dag(\overline{\mathbf{Q}}_0\overline{\mathbf{R}}_0)))||(\mathbf{Q}_{0\backslash\mathbf{S}}\mathbf{R}_{0\backslash\mathbf{S}})^T\mathbf{s}^{l}||_{2,\infty}.
\end{aligned}
\end{equation}

Combining (\ref{impliesbols}), (\ref{lemma485bols}) and Lemma \ref{lemma7}, the condition
\begin{equation}\label{combineyieldsbols}
||(\mathbf{Q}_{0\backslash\mathbf{S}}\mathbf{R}_{0\backslash\mathbf{S}})^T\mathbf{s}^{l}||_{2,\infty}>\frac{2(2-(k-\mathcal{T}_B)d\mu_B)||(\mathbf{Q}\mathbf{R})^T\mathbf{n}^{l}||_{2,\infty}}{2-(k-\mathcal{T}_B)d\mu_B-2\mathcal{T}_Bkd\mu_B}
\end{equation}
ensures that BOLS selects a correct block in the $(l+1)$-th iteration.
Observe that
\begin{equation}\label{notethatbols}
\begin{aligned}
&||(\mathbf{Q}_{0\backslash\mathbf{S}}\mathbf{R}_{0\backslash\mathbf{S}})^T\mathbf{s}^{l}||_{2,\infty}\\
=&||(\mathbf{Q}_{0\backslash\mathbf{S}}\mathbf{R}_{0\backslash\mathbf{S}})^T(\mathbf{I}-\mathbf{P}_{\mathbf{S}^l})\mathbf{D}_{0\backslash\mathbf{S}}\mathbf{x}_{0\backslash\mathbf{S}}||_{2,\infty}\\
\geq&\frac{||(\mathbf{Q}_{0\backslash\mathbf{S}}\mathbf{R}_{0\backslash\mathbf{S}})^T(\mathbf{I}-\mathbf{P}_{\mathbf{S}^l})\mathbf{D}_{0\backslash\mathbf{S}}\mathbf{x}_{0\backslash\mathbf{S}}||_{2}}{\sqrt{k-l}}\\
\geq&\frac{(1-(kd-1)\mu)||\mathbf{x}_{0\backslash\mathbf{S}}||_{2}}{\sqrt{k-l}}.
\end{aligned}
\end{equation}
Combining (\ref{combineyieldsbols}) and (\ref{notethatbols}) yields
\begin{equation}\label{thelastth}
||\mathbf{x}_{0\backslash\mathbf{S}}||_{2}>\frac{2\sqrt{k-l}(2-(k-\mathcal{T}_B)d\mu_B)||(\mathbf{Q}\mathbf{R})^T\mathbf{n}^{l}||_{2,\infty}}{(2-(k-\mathcal{T}_B)d\mu_B-2\mathcal{T}_Bkd\mu_B)(1-(kd-1)\mu)}.
\end{equation}

Finally, using Lemma \ref{lemma8}, the proof is concluded.
\end{IEEEproof}

\section*{Acknowledgment}

This work was supported by the National Natural Science
Foundation of China (61871050) and the NSF grant \#CCF-1527396.

\ifCLASSOPTIONcaptionsoff
  \newpage
\fi



%

\bibliographystyle{IEEEtran}

\end{document}